\documentclass[preprint,12pt]{elsarticle}

\makeatletter
\def\ps@pprintTitle{%
  \let\@oddhead\@empty
  \let\@evenhead\@empty
  \let\@oddfoot\@empty
  \let\@evenfoot\@oddfoot
}
\makeatother

\usepackage{amsmath}
\usepackage{amsfonts}
\usepackage{dsfont}
\usepackage{accents}
\usepackage{subfig}
\usepackage[ruled]{algorithm2e}
\usepackage{xcolor}
\usepackage[normalem]{ulem} 
\usepackage{hyperref}
\hypersetup{
    colorlinks=true,
    linkcolor=blue,
    }

\newcommand\thickbar[1]{\accentset{\rule{.4em}{.8pt}}{#1}}
\newcommand*{\vertbar}{\rule[-1ex]{0.5pt}{2.5ex}}
\newcommand*{\horzbar}{\rule[.5ex]{2.5ex}{0.5pt}}

\journal{ }

\begin{document}

\begin{frontmatter}



\title{Non-intrusive reduced order models for partitioned fluid-structure interactions}


\author[label1]{TIBA Azzeddine}
\author[label2,label5]{DAIRAY Thibault}
\author[label3]{DE VUYST Florian}
\author[label1]{MORTAZAVI Iraj}
\author[label4]{BERRO RAMIREZ Juan-Pedro}

\affiliation[label1]{organization={M2N, CNAM},%Department and Organization
            addressline={2, Rue Conté}, 
            city={Paris},
            postcode={75003}, 
            state={},
            country={France}}

\affiliation[label2]{organization={Manufacture Française des Pneumatiques Michelin},%Department and Organization
            addressline={Place des Carmes-Dechaux}, 
            city={Clermont-Ferrand},
            postcode={63000}, 
            state={},
            country={France}}

\affiliation[label5]{organization={Centre Borelli, CNRS, Université Paris Saclay, ENS Paris Saclay},%Department and Organization
addressline={4, avenue des Sciences}, 
city={Gif-sur-Yvette},
postcode={91190}, 
state={},
country={France}}

\affiliation[label3]{organization={BMBI, UMR 7338, Université de Technologie de Compiègne},%Department and Organization
            addressline={Rue du docteur Schweitzer}, 
            city={Compiègne},
            postcode={60203}, 
            state={},
            country={France}}

\affiliation[label4]{organization={Altair Engineering France},%Department and Organization
            addressline={Rue de la Renaissance}, 
            city={Antony},
            postcode={92160s}, 
            state={},
            country={France}}

\begin{abstract}
The main goal of this work is to develop a data-driven Reduced Order Model (ROM) strategy from high-fidelity simulation result data of a Full Order Model (FOM).
The goal is to predict at lower computational cost the time evolution of  solutions of Fluid-Structure Interaction (FSI) problems.

For some FSI applications, the elastic solid FOM (often chosen as quasi-static) can take far more computational time than the fluid one. In this context, for the sake of performance one could only derive a ROM for the structure and try to achieve a partitioned FOM fluid  solver coupled with a ROM solid one.
In this paper, we present a data-driven partitioned ROM on two study cases: (i) a simplified 1D-1D FSI problem representing an axisymmetric elastic model of an arterial vessel, coupled with an incompressible fluid flow; (ii) an incompressible $2D$ wake flow over a cylinder facing an elastic solid with two flaps.
We evaluate the accuracy and performance of the proposed ROM-FOM strategy on these cases while investigating the effects of the model's hyperparameters. We demonstrate a high prediction accuracy and significant speedup achievements using this strategy.

\end{abstract}

% 85 characters MAX for each bullet point
\begin{highlights}
\item A data-driven reduced order model for partitioned fluid-structure interactions
\item A new approach, coupling a reduced order solid model, and a full order fluid model
\item Reduction of elastic quasi-static solid models coupled with less expensive fluid models
\item Linear and nonlinear dimensionality reduction for forces and displacement fields
\item Parsimonious regression models to learn the solid model in the latent space
\end{highlights}


\end{frontmatter}


\section{Introduction}
\label{}

Fluid-structure interaction (FSI) is the class of mechanical problems dealing with the coupling and interactions between a deformable solid body subject to a fluid loading and a fluid flow. FSI simulations with strong two-way coupling are usually computationally expensive, due to both kinematics and dynamics coupling of the two systems, and the structure of the spatiotemporal dynamics. Although Full Order Models (FOMs) are available and can be discretized using popular numerical methods (\textit{e.g.} finite elements, finite volumes, particle methods ...), the computational cost associated with the simulations is often very high and makes them intractable to predict High-Fidelity (HF) solutions on long-term time periods. In this paper, we will especially focus on simulations based on moving fluid domain methods, the most popular one being the Arbitrary Lagrangian-Eulerian (ALE) method~\cite{donea_arbitrary_1982}.

When solving FSI problems, two main strategies arise, namely the partitioned and monolithic approaches. In the monolithic approach, both the solid and fluid systems are considered as a whole, and the governing equations for both physics are solved at once. While this approach is more robust to the nature of the coupling, due to its abilities to satisfy the coupling conditions exactly\cite{morton_accuracy_1998, bathe_finite_2004, heil_efficient_2004, fernandez_algorithms_2009}, it comes with significant computational and mathematical difficulties, due to the complexity of solving both fluid \textit{and} solid equations simultaneously, while not allowing for the use of well-validated existing structural and fluid solvers.

Partitioned approaches however tackle these challenges with strategies that involve solving the different physics separately, allowing for the use and coupling of available high-fidelity solvers, even in a black-box fashion \cite{uekermann_partitioned_2016, bogaers_quasi-newton_2014}. Specifically, the solid and fluid problems are solved at each time step, and the pressure, velocity and displacements at the interface are communicated in-between to satisfy dynamic, kinematic and geometric coupling conditions respectively. When dealing with situations where the coupling is not very strong, \textit{i.e.} when the effect of one subproblem (\textit{e.g.} solid) on the coupling is significantly less important than the other (\textit{e.g.} fluid), "explicit" schemes, also called "loosely coupled schemes", solve each subproblem only once at each time step, which proved to provide good results in numerous "mildly-coupled" problems (\textit{e.g.} aeroelasticity) \cite{guruswamy_unsteady_1990, piperno_partitioned_1995, farhat_two_2000, lesoinne_geometric_1996}. However, in situations that involve strong fluid-structure coupling, these schemes may  be unstable \cite{nobile_numerical_2001, causin_added-mass_2005, felippa_partitioned_2001}. The coupling constraint needs to be enforced more strongly in an implicit way, involving a fixed-point problem solved using an inner loop of subiterations at each time step \cite{uekermann_partitioned_2016, degroote_performance_2009, degroote_stability_2008, bogaers_quasi-newton_2014}.  This pinpoints the core reason why strongly coupled FSI simulations have a significant computational cost.
%Theses are also called "implicit" schemes.

Reduced Order Models (ROMs) enable efficient computations by reducing large systems, and are now more and more used in industrial applications.
The Proper Orthogonal Decomposition (POD) method is one of the most used ingredients
%popular reduction methods 
in reduced order modeling. The POD method extracts low dimensional linear subspaces from HF data usually obtained with HF simulation results. A ROM can then be built by projecting the FOM equations on the low-order POD basis \cite{sirovich_turbulence_1987, lassila_model_2014, bui_thanh_parametric_2008, veroy_posteriori_nodate}. Projections methods (\textit{e.g.} POD-Galerkin projection) require knowledge of the governing equations. For that reason, they are considered as Physics-based models.
Technically speaking, the projection step requires the access to the source code. The code-intrusive feature of projection-based ROMs can be a shortcoming of their applicability.

Recently, some non-intrusive ROMs have been used in FSI problems using different approaches, for example using linear interpolation of the POD modes and coefficients for parameterized problems \cite{shinde_galerkin-free_2019}, or using Radial-Basis Function (RBF) interpolation of POD coefficients in the context of immersed-shell methods \cite{xiao_non-intrusive_2016}. Hybrid methods combining machine learning surrogates and non-intrusive POD were also used to construct ROMs for FSI \cite{fresca_pod-dl-rom_2022, gupta_hybrid_2022, miyanawala_hybrid_2019, lee_parametric_2023}. Others have also used purely data-driven methods for FSI ROMs (in a weak coupling setting), \citet{zhang_data-driven_2022} for example used convolutional autoencoders and system identification methods to learn the dynamics of each subproblem, while Artificial Neural Networks (ANN) were used for predicting both the solid and fluid solutions in \cite{DL-FSI}.

In this work, we are interested in cases where one solver (\textit{e.g.} solid) has a significantly greater computational cost than the other (\textit{e.g.} fluid). This is for instance the case when nonlinear elastic structural problems under quasi-static loading (inertial effects being neglected) are coupled with incompressible flows in low to medium Reynolds numbers. This suggests the design of a ROM-FOM coupling approach where a structural ROM predicts the response of the FOM solver in a modular fashion, i.e communicating the displacement and/or the velocity at the interface, from the fluid viscous and pressure forces taken as input.
We first find low-dimensional latent spaces in which the forces and displacement fields are embedded, we then learn the relationship between the low-dimensional representations of these fields. This step is done offline in a data-driven manner. The online computations can thus be made non-intrusively and orders of magnitude faster than the FOM computations. Similar approaches for learning data-driven models (in dynamical and static problems) in dimensional latent spaces have been used in -among others- \cite{fries_lasdi_2022, guo_reduced_2018, brunton_discovering_2016}. These works demonstrated the good accuracy and low model complexity achieved when these latent representations are leveraged.

A ROM for a single subdomain coupled with a FOM on the other was done in the independent work of \cite{NNfsiRomFom}. Our work differs from \cite{NNfsiRomFom} in that we first look for latent spaces before finding an adequate regression model between these spaces. Indeed, limited accuracy and speedups were shown in \cite{NNfsiRomFom} even for simple dynamics and a linear elastic structural problem. In fact, it is difficult to train these ANN-based models on such high dimensional spaces while maintaining a high accuracy. Moreover, the Dirichlet conditions are not enforced with a pure ANN solution.
While several works have used non-intrusive ROMs for FSI problems, our proposed approach, along with \cite{NNfsiRomFom} mark -to the best of our knowledge- the only works where a completely data-driven ROM-FOM coupling is explored.

The remainder of this paper is structured as follows: In Sect. \ref{sect:fomfom}, the governing equations of FSI and the partitioned formulation are presented. Then, in Sect. \ref{sect:romfom}, the proposed non-intrusive model reduction approach labeled as ROM-FOM is detailed. The results of the evaluation of this ROM approach in terms of accuracy and stability are presented in Sect. \ref{sect:results}, where two test cases are used as problems on which we use and evaluate the ROM-FOM strategy. Finally, a conclusion is given in Sect. \ref{sect:concl}.

\section{FOM-FOM fluid-structure interaction coupling}\label{sect:fomfom}

A general FSI problem involving an incompressible fluid flow under an ALE description, and an hyperelastic solid can be described by the following equations for each subproblem.
For the fluid subproblem, the incompressible Navier-Stokes in the ALE frame are written as:
    \begin{equation}\label{fluid_eq}
        \begin{cases}
            \rho_f \dfrac{\partial \pmb{v}}{\partial t}_{|\Tilde{\mathcal{A}}} + \rho_f [(\pmb{v} - \pmb{w}).\nabla]\pmb{v} + \nabla p - 2 \,\pmb{\text{div}}(\mu_f \pmb{D}(\pmb{v})) = 0 \quad \text{in}\ \Omega_f(t)\\[1.3ex]
           
            \nabla\cdot\pmb{v}= 0 \quad \text{in }\ \Omega_f(t)\\[1.3ex]

            (2 \mu_f \pmb{D}(\pmb{v}) - p \pmb{I} )\, \pmb{n}_f = \pmb{g}_{N, f}\ \text{in}\ \Gamma_{N, f}(t)\\
        \end{cases}    
    \end{equation}
along with no-slip boundary conditions at walls, imposed velocity profile at the inflow boundaries, and imposed pressure at the outflow boundaries.

For the solid subproblem, the equilibrium and constitutive equations for a static hyperelastic solid are:

        \begin{equation}\label{solid_eq}
        \begin{cases}
             \nabla_{\pmb{X}} \pmb{P} = \pmb{0} \quad \text{in}\; \Omega_{s} \\[1.3ex]
            \pmb{P} = \dfrac{\partial W}{\partial \pmb{F}} \\[1.3ex]
            \pmb{u} = \pmb{0} \; \quad \text{in}\; \Gamma_{D, s}\\[1.3ex]
            \pmb{P} \cdot \pmb{N}_s = \pmb{G}_{N, s}\; \quad \text{in}\; \Gamma_{N, s}\\
        \end{cases}  
        \end{equation}

And finally, the FSI coupling conditions are:
    \begin{equation}\label{coup_eq}
            \begin{cases}
        \pmb{v} = \dfrac{\partial \pmb{u}}{\partial t} = \pmb{w} \quad \text{on}\ \Gamma_{fsi}(t)\\[1.3ex]
        J^{-1} \pmb{F}^T_s \pmb{P} \cdot \pmb{n}_s + (2 \mu_f\, \pmb{D}(\pmb{v}) - p \,\pmb{I} ) \cdot \pmb{n}_f = \pmb{0} \quad \text{on}\ \Gamma_{fsi}(t)\\
        \end{cases}
    \end{equation}

with $\rho_f$ the fluid density, $\mu_f$ the fluid dynamic viscosity, $\pmb{v}$ is the Eulerian fluid velocity and $\pmb{D}(\pmb{v})$ is the fluid strain rate tensor. The fluid equations are described on a moving domain (using the ALE moving frame) $\Omega_f(t)$. The Neumann boundary conditions are defined on the moving boundary $\Gamma_{N, f}(t)$ where $\pmb{n}_f$ represents its exterior normal unit vector.

For the solid problem, the equations are written in the Lagrangian frame with $\nabla_{\pmb{X}}$ the gradient operator in the original configuration, $\pmb{F}_s$ the deformation gradient and $J$ its determinant. $\pmb{P}$ is the first Piola-Kirchoff stress tensor and $\Gamma_{D, s}$ and $\Gamma_{N, s}$ are the Dirichlet and Neumann boundaries respectively, in the original configuration as well, whereas $\pmb{n}_s$ is the normal vector in the \textit{current} configuration.  The vector field $\pmb{u}$ is the solid displacement field and $\pmb{G}_{N, s}$ is the traction force in the original configuration. The material model is described in the stored energy density function~$W$.

The notation $\Tilde{\mathcal{A}}$ represents the ALE mapping from the reference domain (e.g the $t=0$ configuration) to the computational domain and $\pmb{w}$ is the ALE velocity, and $\Gamma_{fsi}$ refers to the "\textit{wet} interface", where coupling between the solid and fluid happens. 

As already mentioned above, in this paper we only consider quasi-static solid conditions meaning that the acceleration term is supposed to be negligible, so we get the elliptic solid problem (\ref{solid_eq}). 
We consider situations where a quasi-static loading is applied, resulting in steady-state nonlinear problems. In this context, the dynamics of the solid do not affect the global FSI problem, and thus are neglected. The solid displacement, however, still affects the strength of the FSI coupling.

In the context of partitioned FSI simulations, we will use the \textit{Dirichlet-Neumann} coupling formulation that allows a 'black-box' FSI coupling. We represent the fluid solver operator as $\mathcal{F}$: 
\begin{equation}
  \mathcal{F}: \mathbb{R}^N \rightarrow \mathbb{R}^N\;;\; \pmb{u}_{|\Gamma_{fsi}} \rightarrow \pmb{f}_{|\Gamma_{fsi}} 
\end{equation}
where $\pmb{u}_{|\Gamma_{fsi}}$ is the displacement field $N$ is the number of interface nodes and $\pmb{f}_{|\Gamma_{fsi}}$ represents the fluid viscous and pressure forces at $\Gamma_{fsi}$:
\begin{equation}
  \pmb{f}_{|\Gamma_{fsi}} = (2 \mu_f \pmb{D}(\pmb{v}) - p \pmb{I} ) \cdot \pmb{n}_{f|\Gamma_{fsi}}.
\end{equation}
Similarly, the solid operator is defined as:
\begin{equation}\label{solid_operator}
  \mathcal{S}: \mathbb{R}^N \rightarrow \mathbb{R}^N\;;\; \pmb{f}_{|\Gamma_{fsi}} \rightarrow \pmb{u}_{|\Gamma_{fsi}}  .
\end{equation}
In fully implicit schemes, the coupling conditions can be enforced  using a fixed-point formulation of the problem (\ref{fluid_eq})-(\ref{coup_eq}):
\begin{equation}\label{dirich_neum}
    (\mathcal{F} \circ \mathcal{S})(\pmb{f}_{|\Gamma_{fsi}}) = \pmb{f}_{|\Gamma_{fsi}}.
\end{equation}
One approach to solve (\ref{dirich_neum})  at each time step is to compute Picard iterations plus a fixed-point acceleration using Quasi-Newton methods for the FSI problem:
\begin{equation}
    (\mathcal{F} \circ \mathcal{S})(\pmb{f}_{|\Gamma_{fsi}}) - \pmb{f}_{|\Gamma_{fsi}} = \pmb{0}
\end{equation}
(see \cite{degroote_performance_2009, haelterman_improving_2016} for more details on the acceleration method used here).

\textbf{Remark}: We note that a "hidden" step consists of mapping the displacement field from the solid mesh to the fluid mesh, and similarly a mapping of the forces from the fluid mesh towards the solid mesh. In fact, our ROM will compute the displacement solution after the mapping of the fluid forces has been performed, so that it gives an output in the same space $\mathbb{R}^N$ of the input.
%
% When dealing with strong FSI coupling, full subiterations are often needed at each time step, 
%
%%%%%%%%%%%%%%%%%%%%%%%%%%%%%%%%%%%%%%%%%%%%%%%%%%%%%%%%%%%%%%%%%%%%%%%%%
% SUBSECTION (13pt, times new roman, bold)
%%%%%%%%%%%%%%%%%%%%%%%%%%%%%%%%%%%%%%%%%%%%%%%%%%%%%%%%%%%%%%%%%%%%%%%%%
\section{Non-intrusive ROM-FOM coupling strategy}\label{sect:romfom}
%--------------------------------------------------------
%
The goal of the ROMs used in this work is to reduce the overall computational cost of the FSI problem through the order reduction of the solid subproblem only. Using partitioned FSI schemes allows for the replacement of the "\textit{module}" of the solid solver~$\mathcal{S}$ with a new ROM solver $\mathcal{S}^{'}$:
\begin{equation}
  \mathcal{S}^{'}: \mathbb{R}^N \rightarrow \mathbb{R}^N\;;\; \pmb{f}_{|\Gamma_{fsi}} \rightarrow \hat{\pmb{u}}_{|\Gamma_{fsi}} 
\end{equation}
and thus achieving a non-intrusive implementation of the model reduction.
In fact, the suggested ROM will also be able to predict the full displacement field (and stress and strain tensor fields) in addition to the interface displacement. But note that only the interaction variables located at the FSI interface are needed to advance the FSI solution in time. The fluid solver, as well as the other components of the FSI algorithm (i.e implicit coupling, Quasi-Newton acceleration ...) remain the same. This produces a non-intrusive ROM-FOM coupling scheme, with an expected reduced computational cost compared to the original FOM-FOM coupling.

This approach has the advantage of minimal dynamics-associated errors. Since the structural model is quasi-static, the high fidelity fluid FOM will handle the dynamics of the complete FSI model, and will work as a kind of corrector of the phase and frequency errors, even if the errors of structural model accumulate with time. Moreover, only solids with path-independent material behaviors are considered. Thus, the structural ROM does not need to be "aware" of the history of the load. This is specifically the case for elastic (that can be nonlinear) materials.

It is worth mentioning that this strategy can particularly achieve significant speedups when the solid FOM is much more expensive than the fluid FOM. Moreover, we assume that in online-computations, the average number of subiterations does not increase compared to the FOM-FOM problem. As we will see in the numerical experiments, this is the case when the solid ROM is accurate enough and stable compared to the FOM.

\subsection{Detailed ROM-FOM methodology}
Since we are interested in solid problems with quasi-static behaviour only, we argue that the solid ROM can ignore the dynamics effects and only take into account the fluid loading at the interface. The HF solution from the FOM-FOM coupling is first used to train our model. Accordingly, two snapshot matrices are created from the forces at the interface $\pmb{F}$ and the full solid displacement field $\pmb{U}$, collecting $m$ snapshot solutions from all the subiterations at each time step during the FOM-FOM computation.
We note that the force field is discretised on the solid mesh interface, meaning that we collect the force data after the mesh mapping step during the FSI solution schemes.

The suggested ROM will perform (i) a dimensionality reduction of the input $\pmb{f}_{|\Gamma_{fsi}}$ and output $\hat{\pmb{u}}$ of the solver, (ii) solve a regression problem in the low-dimensional latent space and (iii) reconstruct the displacement field in the original physical space.

\textbf{Dimensionality reduction}:
 Specifically, we use the Principal Component Analysis (PCA) method (also called POD in this context) to find the best linear subspace of rank $r_f$ on which the forces field is projected, we will refer to the POD modes as $\pmb{\Phi}_f$ $\in \mathbb{R}^{N \times r_f}$. This means that the force field can be written as: 

 \begin{equation}\label{reduc}
     \pmb{f}_{|\Gamma_{fsi}}(t) = \sum_i^{r_f} \pmb{\Phi}_{fi} \Tilde{\pmb{f}}_i(t)
 \end{equation}
meaning that the snapshot matrix of the force field $\pmb{F} \in \mathbb{R}^{N \times m}$ can be written like:
 \begin{equation}
    \pmb{F} = \pmb{\Phi}_f \Tilde{\pmb{f}}
 \end{equation}
 where $\Tilde{\pmb{f}} \in \mathbb{R}^{r_f \times m}$ are the coordinates of the forces snapshots in the reduced POD subspace. The discrete POD modes can be obtained in the offline phase using a Singular Value Decomposition (SVD) of the snapshot matrix: 
 \begin{equation}
    \pmb{F} = \pmb{\Phi}_f \pmb{\Sigma}_f \pmb{V}_f^*
 \end{equation}
 
Accordingly, the reduced coordinates $\Tilde{\pmb{f}}$ can be found using an orthogonal projection on the POD modes
 \begin{equation} \label{proj}
    \Tilde{\pmb{f}} = \pmb{\Phi}_f^{T} \pmb{F}
 \end{equation}

%and $\pmb{\Phi}_u$ $\in \mathbb{R}^{N_s \times r_u}$ respectively  where $N_s$ is the number of displacement degrees of freedom in the solid domain.
On the other hand, a dimensionality reduction method is also applied on the displacement field. This time, however, a retrieval of the predicted full displacement field will be needed, since a decoding from the latent space of the displacement is the ultimate step in our approach. Thus, the accuracy of the full displacement reconstruction is crucial, and the error associated with its reduction needs to be as low as possible. For these reasons, we use a quadratic manifold representation of the displacement field \cite{geelen_operator_2023}. Using the method introduced in \cite{geelen_operator_2023}, the residual error of the approximation (\ref{reduc}) is modeled using the quadratic terms of the reduced coordinates, keeping $r_u$ modes: 

 \begin{equation}
     \hat{\pmb{u}}(\Tilde{\pmb{f}}) = \sum_i^{r_u} \pmb{\Phi}_{ui} \Tilde{\pmb{u}}_i(\Tilde{\pmb{f}}) + \sum_j^{\frac{1}{2}r_u (r_u+1)} \thickbar{\pmb{\Phi}}_{uj} (\Tilde{\pmb{u}}(\Tilde{\pmb{f}})  \otimes \Tilde{\pmb{u}}(\Tilde{\pmb{f}}))_j
 \end{equation}

where $\otimes$ is the Kronecker product (resulting in all the polynomial terms with the exception of the linear terms) of the reduced forces coordinates $\Tilde{\pmb{u}} \in \mathbb{R}^{r_u \times m}$. For the collected displacement snapshots $\pmb{U}$, we write: 
\begin{equation}\label{quad_recons}
    \pmb{U}  =  \pmb{\Phi}_{u} \Tilde{\pmb{u}} +   \thickbar{\pmb{\Phi}}_{u} \pmb{X} 
\end{equation}
where $\pmb{X} =         \left[
  \begin{array}{cccc}
        \vertbar & \vertbar &        & \vertbar \\
         \Tilde{\pmb{u}}_{1} \otimes  \Tilde{\pmb{u}}_{1}   & \Tilde{\pmb{u}}_{2} \otimes  \Tilde{\pmb{u}}_{2}    & \ldots & \Tilde{\pmb{u}}_{m} \otimes  \Tilde{\pmb{u}}_{m}    \\
        \vertbar & \vertbar &        & \vertbar 
      \end{array}     \right]
 \in \mathbb{R}^{\frac{1}{2}r_u (r_u+1) \times m}$
 
Similarly, at the online stage, at any given iteration, the full displacement field is:
\begin{equation}\label{current_recons}
    \hat{\pmb{u}}_{current}  =  \pmb{\Phi}_{u} \Tilde{\pmb{u}}_{current} +  \thickbar{\pmb{\Phi}}_{u} \pmb{X}_{current} 
\end{equation}

The columns in the quadratic mapping operator $\thickbar{\pmb{\Phi}}_{u}$ $ \in \mathbb{R}^{N_u \times \frac{1}{2}r_u (r_u+1)}$ represent the quadratic modes, which are orthogonal to the POD subspace $\pmb{\Phi}_{u}^T \thickbar{\pmb{\Phi}}_{u} = \pmb{0}$, hence modeling correctly the error term not accounted for in a linear manifold approximation.

The quadratic mapping matrix can also be obtained in a data-driven fashion during the offline phase. In fact, after a first step of learning the POD modes $\pmb{\Phi}_{u}$, a second step consists of a linear least squares problem:
\begin{equation}\label{vbar}
    \thickbar{\pmb{\Phi}}_{u} =  arg \min_{\thickbar{\pmb{\Phi}}  \in \mathbb{R}^{N_u \times \frac{1}{2}r_u (r_u+1)} } \frac{1}{2} ||(\pmb{I} - \pmb{\Phi} \pmb{\Phi}^T) \pmb{U}  -  \thickbar{\pmb{\Phi}} \pmb X   ||^2_F
\end{equation}

Regarding the choice of the number of modes $r_u$, different strategies can be used. The most usual one being an energy-based criterion: using the singular values $\sigma_i$ from the SVD of the snapshot matrix, a certain threshold $\epsilon$ representing the percentage of the energy contained in theses snapshots can be chosen, and the number of modes can be determined as:

\begin{equation}\label{energy_sing}
\begin{aligned}
\min_{r_u\in[1, d]} \quad & S = \frac{\sum_i^{r_u} \sigma_i^2}{\sum_i^d \sigma_i^2}\\
\textrm{s.t.} \quad &  S \leq \varepsilon\\
\end{aligned}
\end{equation}

where $d$ is the minimum dimension of the snapshot matrix. However, this criterion is not always the best option. For example, in problems with solution of a slow decaying Kolmogorov width, a large number of modes is generally needed to have a sufficiently accurate linear approximating subspace (see for example \cite{lassila_model_2014}). To this end, we suggest a cross-validation strategy, where a portion of the snapshot data can be used for testing, and where an increasing number of modes is tested until a minimum ( or a plateau ) of the overall testing error is reached. We suggest that this should be preferred to an energy-based criterion whenever possible (and computationally feasible). We also note that snapshot data should be scaled by removing the mean field in order to retrieve the POD subspace correctly. 

\textbf{Remark:} We note that in \cite{geelen_operator_2023}, the authors also present an additional step of column selection on $\thickbar{\pmb{\Phi}}$ through another optimization problem, and thus minimizing the number of quadratic terms used in the quadratic manifold. In our work, we bypass this step and keep all the quadratic terms, since this choice generally provides the best reconstruction accuracy, and since in the usual problems we encounter, the solid displacement field is sufficiently smooth so that only few modes are needed, meaning that selecting fewer quadratic columns will only result in a very slight performance gain.

To summarize this step, our ROM approach in the online stage computes an encoding of the fluid forces field at each iteration, using a linear projection on the POD subspace, thus only using the projection part through (\ref{proj}). For the ROM output, we use quadratic manifolds for the displacement field, this time only using the decoder component, via the reconstruction (\ref{quad_recons}).

\textbf{Remark 2:}
From a Dirichlet-Neumann formulation perspective, only the displacement values at the interface is required from the solid operator $\mathcal{S}$ (and its ROM alternative $\mathcal{S}'$) as described in (\ref{solid_operator}). As a result, and unlike the solid FOM solver that computes the full displacement field at the solid domain, the ROM solver should - at any given iteration - be able to exchange the displacement values at the interface only. This can be done in a straightforward manner when using a decoder based on a reconstruction on a POD basis. If we construct a mapping matrix from the solid nodes to the interface nodes:
\begin{equation}
    \pmb{K} =  \left[
  \begin{array}{ccc}
        \horzbar & \pmb{\mathds{1}}_1 &   \horzbar   \\
        \horzbar & \pmb{\mathds{1}}_2 &   \horzbar   \\
        \vdots & \vdots &   \vdots   \\
        \horzbar & \pmb{\mathds{1}}_N &   \horzbar   \\
      \end{array}     \right] \in \mathbb{R}^{N \times N_u}
\end{equation}
where $\pmb{\mathds{1}}_i = \left[0\ldots 1 \ldots  0 \right]$ are zero row vectors with a $1$-valued element at the $j$th column where $j$ is the index of the solid mesh node corresponding to the interface node $i$. 
We can then compute the matrix multiplications $\pmb{K} \pmb{\Phi}_u$ and $\pmb{K} \thickbar{\pmb{\Phi}}_u$ during the offline stage and store the resulting matrices. At each iteration, the solid ROM operator can thus compute the displacement values at the interface directly, replacing (\ref{current_recons}) with:
\begin{equation}\label{new_reconstruct}
    \hat{\pmb{u}}_{current\;|\Gamma_{fsi}}  =  \pmb{K}\pmb{\Phi}_{u} \Tilde{\pmb{u}}_{current} +  \pmb{K}\thickbar{\pmb{\Phi}}_{u} \pmb{X}_{current} 
\end{equation}
This is particularly important since the number of interface nodes is usually much smaller than the whole solid mesh nodes $N \ll N_u$ and (\ref{new_reconstruct}) will be faster to compute than (\ref{current_recons}). In addition, since there is not much use to retrieve the full displacement field at the non-convergent coupling iterations, the ROM can store the reduced displacement coordinates $\Tilde{\pmb{u}}_{current}$ only at the convergent iterations and compute the reconstruction (\ref{quad_recons}) at the end of the simulation, getting back the full displacement field, and enabling the computation of the strain and stress fields for example.

\textbf{The regression problem} on the other hand can be solved using different existing methods with a regression operator
\begin{equation}
    \mathcal{I}: \mathbb{R}^{r_f} \rightarrow \mathbb{R}^{r_u}\;;\; \Tilde{\pmb{f}}  \rightarrow  \Tilde{\pmb{u}}
\end{equation}
where $\Tilde{\pmb{f}} = \pmb{\Phi}_f^T  \pmb{f}_{|\Gamma_{fsi}}$ and $\Tilde{\pmb{u}} = \pmb{\Phi}_u^T \pmb{u}$ are the coordinates of the force field and the displacement field in the reduced bases respectively.
In our experiments, the regression methods that provided the best accuracy are reduced basis function (RBF) interpolation \cite{wahba_spline_1990} and low-degree polynomial sparse approximation. \medskip

The proposed ROM algorithms in the offline and online stages are summarized in Algorithms \ref{alg:offline} and \ref{alg:online} below, and in the illustration in Figure \ref{fig:romFOMillus}.

\SetKwComment{Comment}{/* }{ */}
\SetKwInput{KwData}{Input}

\begin{algorithm}
\caption{ROM-FOM - Solid ROM Offline stage}\label{alg:offline}
\KwData{$\pmb{F}$ the force snapshot matrix, $\pmb{U}$ the displacement snaphot matrix, $r_f$ the number of selected force modes, $r_u$ the number of selected displacement modes, $\pmb{K}$ the mapping matrix from the solid to the interface nodes}
\KwResult{$\{\pmb{\Phi}_f$, $\pmb{\Phi}_u$, $\mathcal{I}(\cdot)$$\}$}
\vline

Compute $r_f$ left singular vectors $\pmb{\Phi}_f$ from the SVD of $\pmb{F}$, compute $\Tilde{\pmb{f}} = \pmb{\Phi}_f^T  \pmb{F}$ \;
\vline

Compute $r_u$ left singular vectors $\pmb{\Phi}_u$ from the SVD of $\pmb{U}$, compute  $\thickbar{\pmb{\Phi}}_{u}$ by solving (\ref{vbar}), compute $\Tilde{\pmb{u}} = \pmb{\Phi}_u^T  \pmb{U}$ \;
\vline

Compute the matrices $\pmb{A} = \pmb{K} \pmb{\Phi}_{u}$ and $\pmb{B} = \pmb{K} \thickbar{\pmb{\Phi}}_{u}$ \;
\vline

Determine the regression operator $\mathcal{I}(\Tilde{\pmb{f}}) \approx \Tilde{\pmb{u}}$ \;

\end{algorithm}
\begin{algorithm}
 \RestyleAlgo{ruled}
\caption{ROM-FOM - Online stage (FSI interaction variables)}\label{alg:online}
\KwData{Current interface force quantity, $\pmb{f}_{|\Gamma_{fsi}, current}$}
\KwResult{Current solid displacement $\hat{\pmb{u}}_{current}$, Current interface displacement $\hat{\pmb{u}}_{|\Gamma_{fsi, current}}$}

Project the fluid loading on the reduced basis: $\Tilde{\pmb{f}}_{current} = \pmb{\Phi}_f^T  \pmb{f}_{|\Gamma_{fsi}, current}$\;
\vline

Predict the new reduced displacement: $\Tilde{\pmb{u}}_{current} = \mathcal{I}(\Tilde{\pmb{f}}_{current}) $\;
\vline

Arrange the quadratic terms of $\Tilde{\pmb{u}}_{current}$  in $\pmb{X}_{current}$\;
\vline

Compute the interface displacement field: $\hat{\pmb{u}}_{current\;|\Gamma_{fsi}}  =  \pmb{A} \Tilde{\pmb{u}}_{current} +  \pmb{B} \pmb{X}_{current}$\;
\vline

  \uIf{coupling convergence}{
    Store  $\hat{\pmb{u}}_{current\;|\Gamma_{fsi}}$ in $\Tilde{\pmb{u}}_{stored}$
  }
\vline

  \uIf{end of simulation}{
    Reconstruct the full displacements     $\pmb{U}  =  \pmb{\Phi}_{u} \Tilde{\pmb{u}}_{stored} +   \thickbar{\pmb{\Phi}}_{u} \pmb{X}_{stored}$
  }

\end{algorithm}
\begin{figure}
    \centering
    \includegraphics[width=1\textwidth]{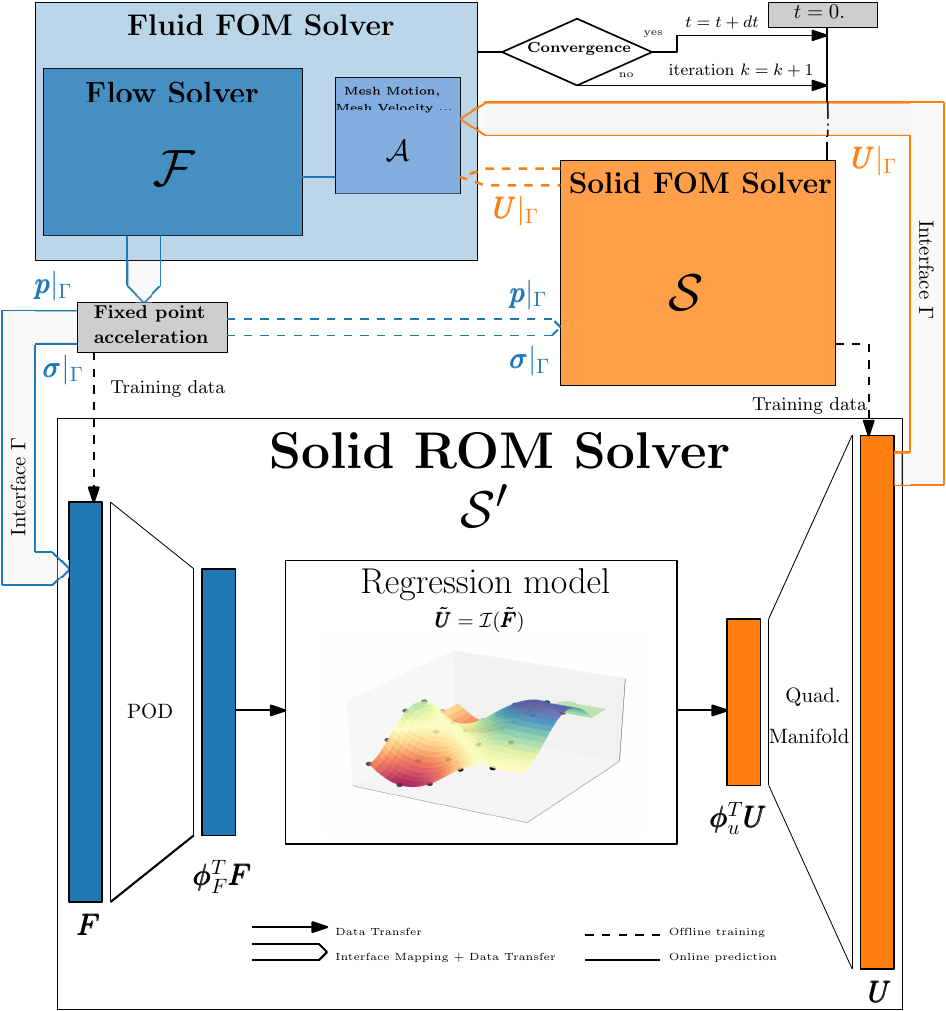}
    \caption{Summarizing illustration of the ROM-FOM approach}
    \label{fig:romFOMillus}
\end{figure}
\subsection{Expected speedups}
In this short section, we give an idea of the overall speedup (denoted by $s$) of the partitioned FSI coupling if the solid ROM solver returns a speedup $\sigma$ compared to the solid FOM solver. As mentioned above we will assume that the number of fixed-point subiterations does not vary between FOM-FOM and ROM-FOM strategies. Let us denote by $T_f$ (resp. $T_s$) the mean computational time taken by the FOM fluid (resp. solid) solver during one time iteration of the FSI coupling. The total FOM-FOM time over a time step is $T_f+T_s$ while the ROM-FOM time is $T_f+T_s/\sigma$. The ROM-FOM speedup is then
\[
s = \frac{T_f+T_s}{T_f+\dfrac{T_s}{\sigma}} = \frac{1+ T_s/T_f}{1+\dfrac{1}{\sigma} T_s/T_f}.
\]
Assume that $T_s/T_f\gg 1$. Then we get the speedup estimation
\begin{equation}
s \approx \frac{T_s/T_f}{1+\dfrac{1}{\sigma} T_s/T_f}
= \frac{\sigma}{1+\sigma \dfrac{T_f}{T_s}}.
\label{eq:speedup}
\end{equation}
Equation~\eqref{eq:speedup} shows that a 'good' solid ROM speedup should be
of the order $T_s/T_f$. Assume for example that $\sigma=T_s/T_f$, then one finds $s = \sigma/2$ and the efficiency of the ROM-FOM FSI strategy is 1/2. More generally, if the solid ROM achieves a solid 
speedup $\sigma=\alpha \dfrac{T_s}{T_f}$ with $\alpha>0$, then
\[
s\approx \left(1-\frac{1}{1+\alpha}\right)\,\frac{T_s}{T_f}.
\]
In particular, the ratio $\dfrac{T_s}{T_f}$ is an upper bound of ROM-FOM FSI speedup.

% \begin{figure}[!htbt]
% \begin{center}
% \setlength{\unitlength}{1mm}
% \begin{picture}(110, 100)
% {\framebox{\epsfig{file=Images/iters_tube, scale=0.48,}}}
% \end{picture}
% \end{center}
% \caption{iters tube.}
% \end{figure}

\section{ROM-FOM coupling assessment for low cost FSI simulations}\label{sect:results}

\subsection{Model evaluation strategy}\label{eval_st} In order to evaluate the proposed ROM approach, for multiple parameter values in a training set $\mathcal{P}$, the results of the ROM-FOM simulation will be tested in a future time prediction. In other words, for a given parameter $\pmb{\mu} \in \mathcal{P} \subset \mathcal{D}$, where $\mathcal{D}$ is the convex hull of $\mathcal{P}$, a FOM-FOM simulation will be performed for a specific time period $t \in [0, T]$. After training the ROM on the obtained high fidelity results (including the coupling subiterations), the structural ROM model can be used for $t \in [T, T_f]$. In addition, the results associated with all the seen parameters can be used together to train a single ROM, and predict solutions for parameters not seen in training $\hat{\pmb{\mu}} \not\in \mathcal{P}$, and for future time regions $t \in [0, T_f]$. In that case, we will also assess the methodology for testing parameters outside the parameter space $\hat{\pmb{\mu}} \not\in \mathcal{D}$. In this work, only parameters associated with the fluid problem are considered. We should note that in the case of time-parameter prediction cases, and in order to avoid ROM-FOM instabilities further down the simulation, the solid FOM needs to be used at some of the first time increments, e.g until a time instant that we will call $T_i$, the purpose of this is that on transient cases, the results in the first timesteps are very different than the rest of the simulation, meaning that the training data is much poorer around these values, this results in lower accuracy of the solid ROM, hence it is best to avoid using the ROM in the first transient timesteps. An illustration of this model evaluation strategy applied on the second example can be seen in Figure \ref{eval_illustr}. Simulation data, used for training and testing are available at the Github repositories (\href{https://github.com/FsiROM/ArterialWallROM/tree/JFS}{ArterialWallROM}) and (\href{https://github.com/FsiROM/DoubleFlap/tree/JFS}{DoubleFlap}) corresponding to the first and second examples respectively.

\subsection{Example 1: 1D toy problem of an elastic arterial vessel model}

\begin{figure}[t]
\begin{center}
\includegraphics[width=.6\textwidth]{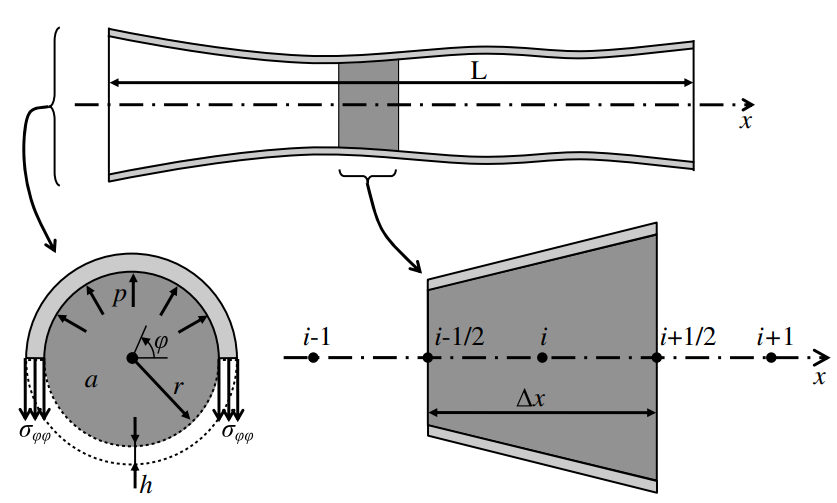}
\end{center}
\caption{Arterial vessel test case schematic explanation (from \cite{degroote_stability_2008}). $\sigma_{\varphi \varphi}$ is the vessel hoop stress, $h$ is the thickness and $\Delta x $ is the length of the finite volume cell.}
\label{testcase1}
\end{figure}

The model of flexible tube and related HF partitioned solvers proposed by \citet{degroote_stability_2008} are used here. The flow is assumed to be incompressible with constant density $\rho$. Both fluid mass and momentum conservation equations (neglecting viscosity) read
\begin{equation}\label{fluid_flow}
    \begin{cases}
	 \partial_t a + \partial_x(a v) = 0, \\
        \partial_t(av)+ \partial_x(a v^2) + \dfrac{a}{\rho}\,\partial_x p = 0,
	\quad t>0, \ x\in [0,L]
 \end{cases}
\end{equation}
where $v$ is the bulk velocity, $a$ is the tube cross section and
$p$ is the pressure. From the fluid side, the unknowns are both
velocity and pressure.
For the solid flexible tube, a quasi-static model 
\[
a = a(p)
\] 
is used (retaining only the vessel stress in the circumferential direction). The following nonlinear elastic stress-strain law is used:
\begin{equation}\label{strs_strain_law}
    \begin{cases}
	 \sigma_{\varphi \varphi} = 12500\, \epsilon_{\varphi \varphi}\quad \text{if} \ |\epsilon_{\varphi \varphi}|< \epsilon_0 \\
   \sigma_{\varphi \varphi} = 2500\, \epsilon_{\varphi \varphi} + 20\quad \text{if} \ \epsilon_{\varphi \varphi}\geq \epsilon_0 \\
   \sigma_{\varphi \varphi} = 2500\, \epsilon_{\varphi \varphi} - 20\quad \text{if} \ \epsilon_{\varphi \varphi}\leq -\epsilon_0 \\
 \end{cases}
\end{equation}
with $\epsilon_0=2\,\, 10^{-3}$.
Figure \ref{testcase1} shows a schematic explanation of this problem.
For further simplification, a zero pressure condition is applied on the right boundary (note that a non-reflective boundary condition was used in~\cite{degroote_stability_2008}).
The prescribed inlet (left face) velocity is computed using the solution of a nonlinear Duffing equation in order to produce a signal with rather complex dynamics. 
\begin{equation}
    \begin{cases}
	 \ddot u(t) = a\, u(t) + b\, u(t)^2 + c\, u(t)^3 + d + p\,cos(f t) + e\, \dot u(t)\,\,\,\, \forall t \in [0, 120] \\
   u(0) = 10\,\,\,;\,\,\,\dot u (0) = 0. \\
   v_{inlet}(t) = g u(t) + h \\
 \end{cases}
\end{equation}
We fix $(a,\ b\ ,c\ ,d\ ,e\ ,g\ , p) = (-1\,\,\,,0\,\,\, ,-0.002\,\,\, ,-1\,\,\, ,-0.02\,\,\, ,1/60\,\,\, ,360)$ and we parameterize this signal with the parameter vector $\pmb{\mu} =(f\,\,\, ,h )^T$ allowing the generation of different frequencies and amplitudes.

\begin{figure}
\begin{center}
% \setlength{\unitlength}{1mm}
% \begin{picture}(210, 80)\label{fig:inlets}
% {\framebox{\epsfig{file=Images/Inlet_v, scale=0.41,}}}
% \end{picture}
\includegraphics[width=1\textwidth]{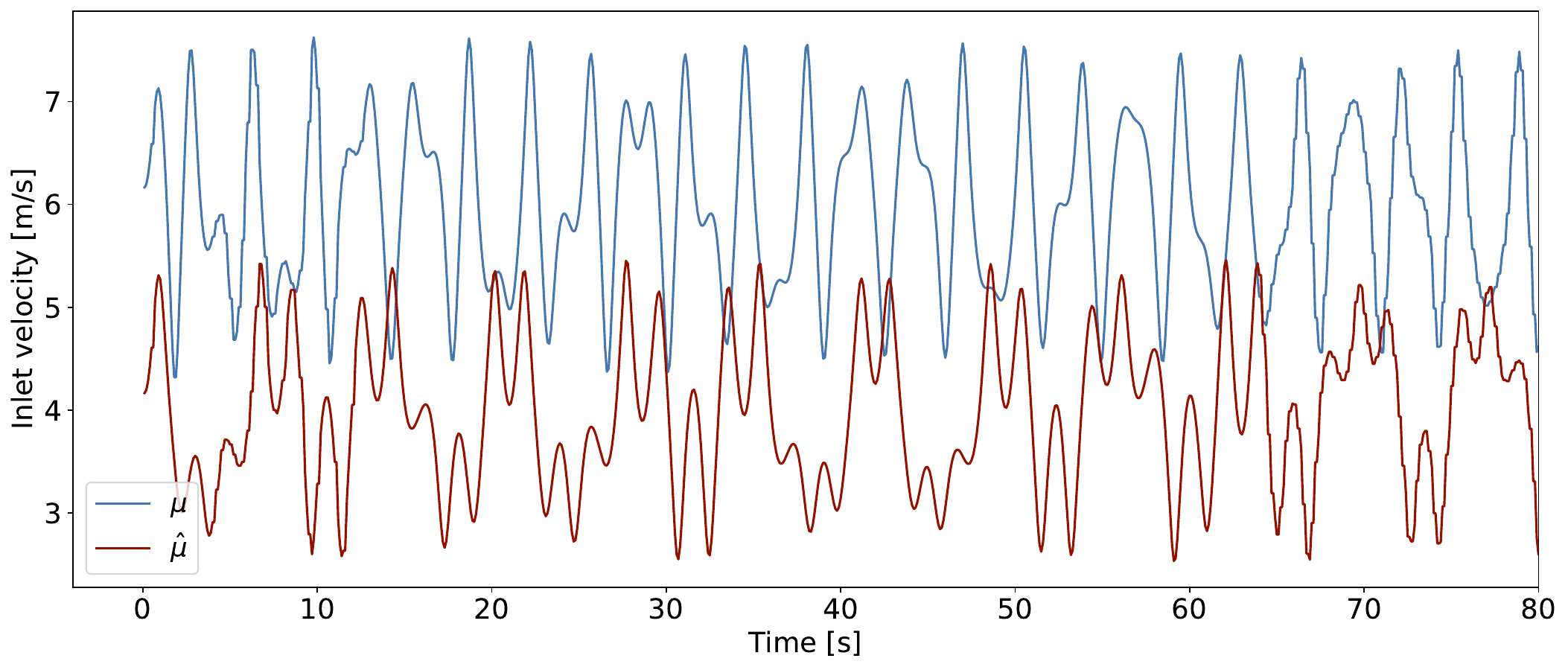}
\end{center}
\caption{Inlet velocity signals corresponding to $\pmb{\mu}$ and $\hat{\pmb{\mu}}$ parameters values.}
\label{inlet_v}
\end{figure}

In order to train the ROM model, a FOM-FOM computation is done on a single inlet velocity case corresponding to $\pmb{\mu}_1 =  (2\,\,\,,6)^T$ (see Figure \ref{inlet_v}), for a simulation time period ending in $T = 18$~s. The fluid flow equations (\ref{fluid_flow}) are solved using a second order finite volume scheme with $100$ cells and the solid section $a(p)$ is computed at each iteration as the solution of a scalar minimization problem. We use the software library \texttt{preCICE} \cite{noauthor_precice_nodate} as a coupling interface for the simulations.

For the considered training time region and including all the coupling subiterations, the data is assembled as 737 snapshots ($m = 737$ and $N = N_u = 101$), and $\{ \pmb{U}, \pmb{F}\} \subset \mathbb{R}^{101 \times 737}$.

\begin{figure}
    \begin{center}
    \includegraphics[width=.55\textwidth]{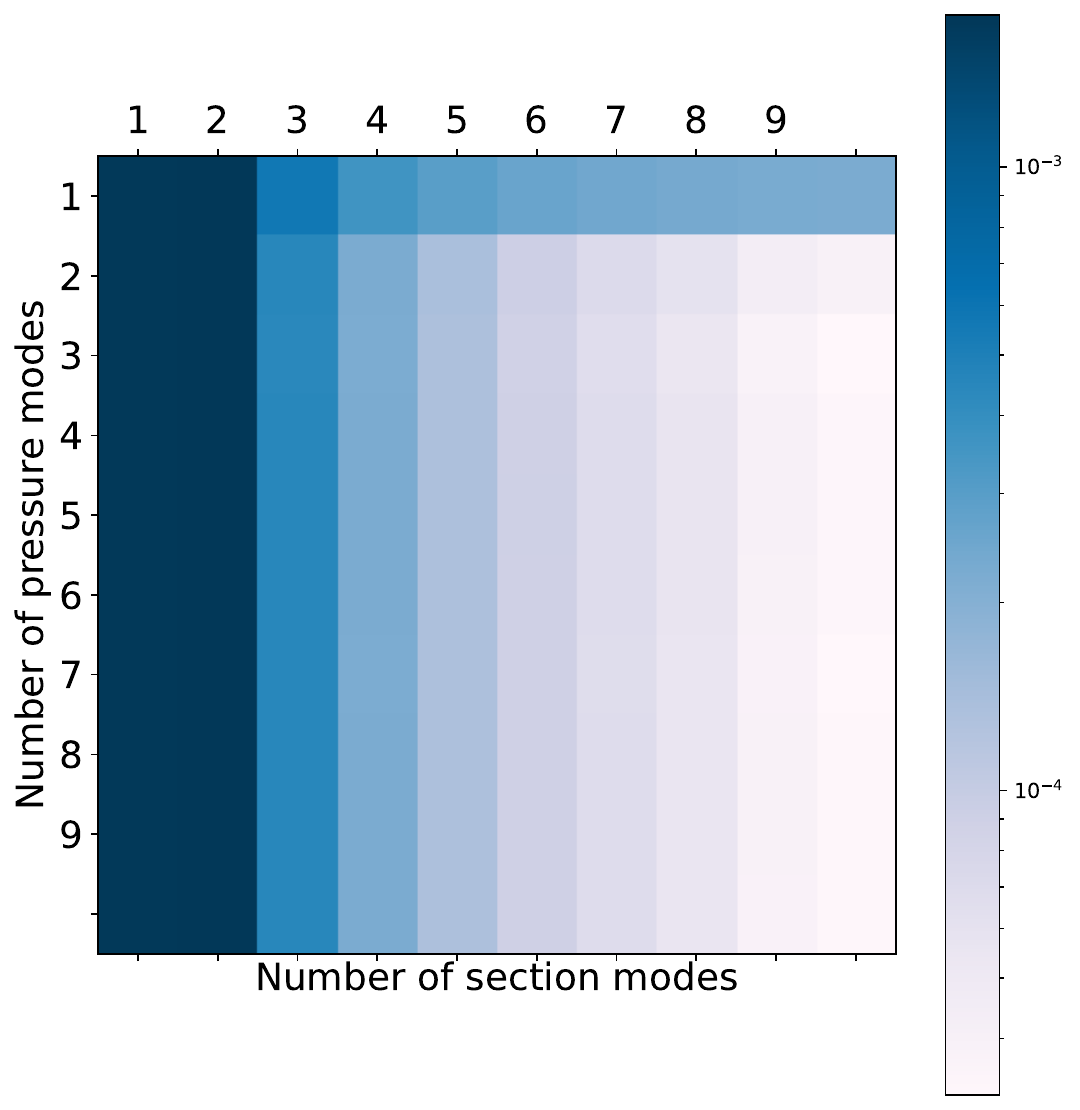}
    \end{center}
    \caption{Validation error $\epsilon$ depending on the number of modes choices for both the pressure and the section fields.}
    \label{cas1_error_map}
\end{figure}

\begin{figure}
    \begin{center}
    \includegraphics[width=.65\textwidth]{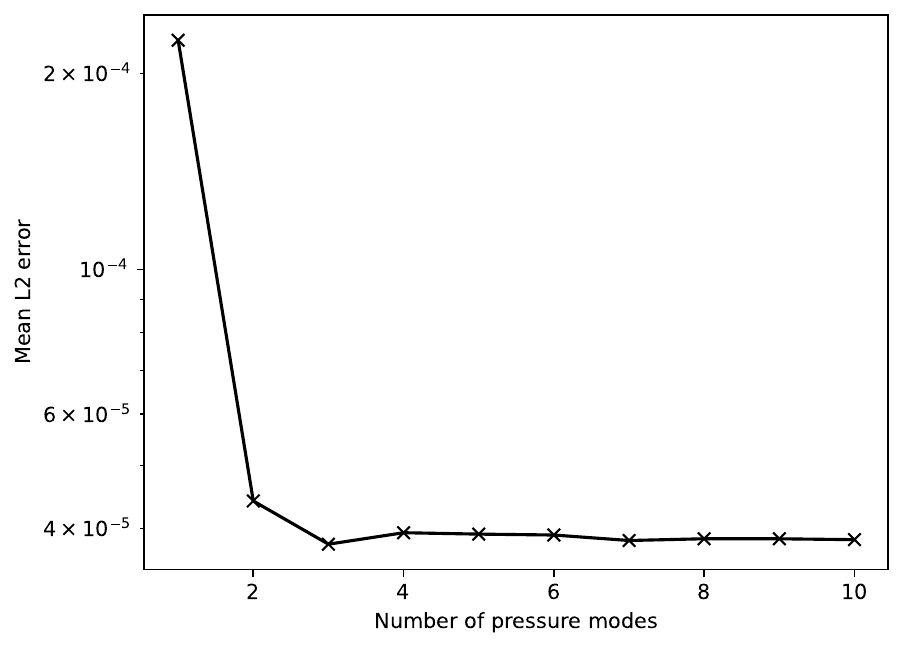}
    \end{center}
    \caption{Validation error $\epsilon$ when using 9 section modes depending on the number of pressure modes.}
    \label{fixed_disp}
\end{figure}

For this problem, the pressure and section fields are very smooth and only very few modes can model the problem solution. In fact, a $99.99\%$ energy criteria like in (\ref{energy_sing}) implies that only one pressure mode and 5 section modes are needed. Nonetheless, we will use cross-validation to determine an optimal number of modes. A random portion making $20\%$ of the training snapshots will be used as testing samples (denoted as $\pmb{F}_{test}$ and $\pmb{U}_{test}$) and while varying the number of modes of the pressure and section fields, the mean of the $L_2$ error on the final output (i.e the section field) will be evaluated upon these samples. In Figure \ref{cas1_error_map}, we show a map of the error $\epsilon$ depending on the number of modes, where 
\begin{equation}
    \epsilon = \frac{1}{m_{test}} \sum_i^{m_{test}} ||\pmb{U}_{test,i} - ROM(\pmb{F}_{test,i})||_2
\end{equation}
$m_{test}$ is the number of testing samples and $ROM(\cdot)$ denotes the final output of the trained ROM. Using theses results, we can see that the test error levels off when we reach 9 modes for the section field and 3 modes for the pressure. More precisely, for 9 section modes, we show the evolution of the error $\epsilon$ for different numbers of pressure modes in Figure \ref{fixed_disp}. This also reinforces the idea that this type of cross-validation should be used whenever possible to determine these hyperparameters, since the use of the energy-based criterion only would result in larger error (as seen in Figure \ref{cas1_error_map}, the accuracy when using 1 pressure mode and 5 section modes is suboptimal). We note that the displacement reconstruction used here only considers the linear POD terms. In fact, this example being a toy example with the goal of predicting solutions with minimal training data, the training snapshots are not rich enough to accurately learn the quadratic operator $\thickbar{\pmb{\Phi}}$.

%(in this case, the section longitudinal field) 
Regarding the regression method, a thin plate spline kernel RBF interpolator is used \cite{wood_thin_2003} with

\begin{equation}
    \mathcal{I}(\Tilde{\pmb{f}}) = \sum_i^m w_i\, \phi(||\Tilde{\pmb{f}} - \Tilde{\pmb{f}}_i||)
\end{equation}
where 
\begin{equation}
    \phi( \pmb{x} ) = \pmb{x}^2 \log(\pmb{x})
\end{equation}
and $\Tilde{\pmb{f}}_i$ are the RBF centers, chosen as the training points of the reduced forces, resulting eventually in a linear system to be solved for the RBF weights $w_i$.

\begin{figure}[!ht]
\begin{center}
% \setlength{\unitlength}{1mm}
% \begin{picture}(210, 90)
% {\framebox{\epsfig{file=Images/tube_pred, scale=0.41,}}}
% \end{picture}
\includegraphics[width=1\textwidth]{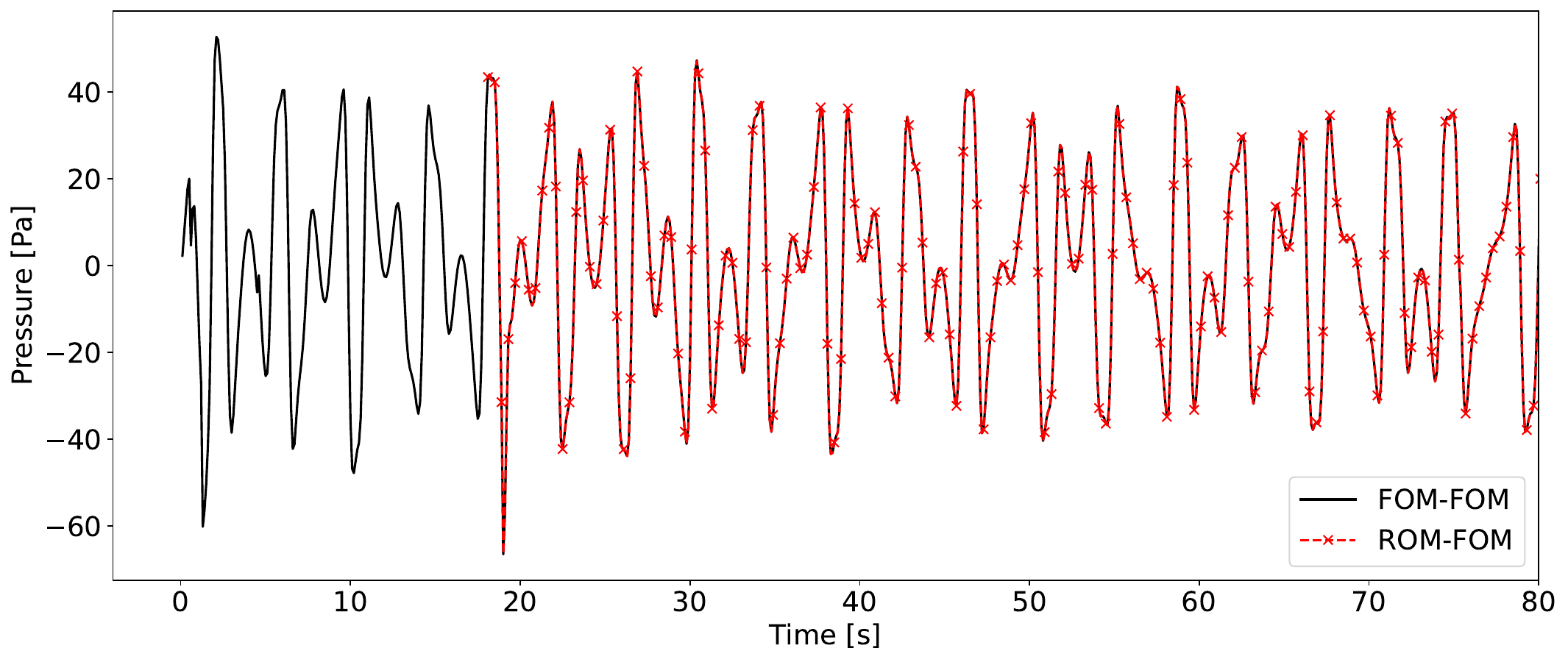}
\end{center}
\caption{The inlet pressure solution using the FOM-FOM (green dashed line with cross marks) and the ROM-FOM (blue solid line). Prescribed inlet velocity corresponding to $\mu_1$. Both training and prediction regimes are depicted. The vertical black line indicates the end of the training time period.}
\label{test1}
\end{figure}

\begin{figure}[!ht]
\begin{center}
% \setlength{\unitlength}{1mm}
% \begin{picture}(210, 90)
% {\framebox{\epsfig{file=Images/tube_mu, scale=0.41,}}}
% \end{picture}
\includegraphics[width=1\textwidth]{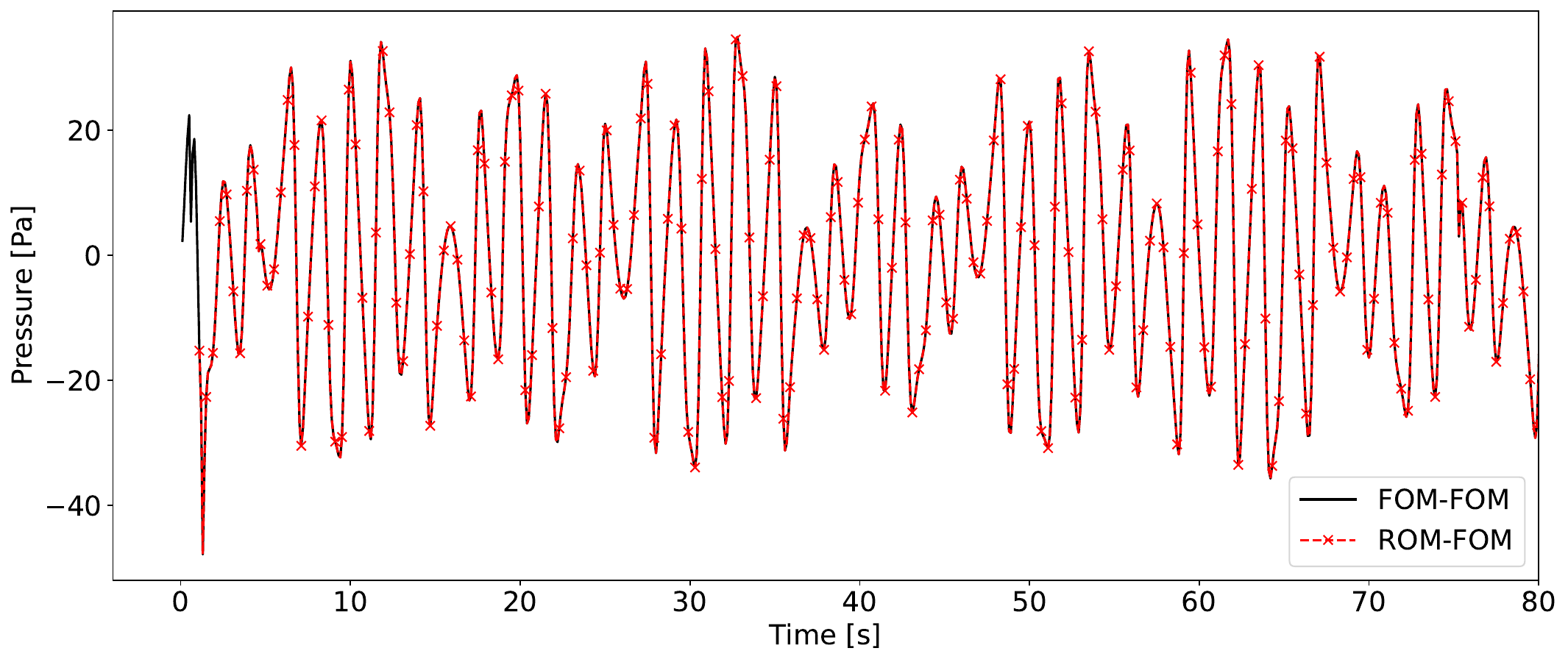}
\end{center}
\caption{The inlet pressure solution using the FOM-FOM and the ROM-FOM. Prescribed inlet velocity corresponds to $\mu_2$. Prediction regime.}
\label{test2}
\end{figure}

\begin{figure}[!ht]
\begin{center}
% \setlength{\unitlength}{1mm}
% \begin{picture}(100, 80)
% \put(   0,  0){\framebox{\epsfig{file=Images/mAc_strstrn, scale=0.5,}}}
% \end{picture}
\includegraphics[width=.6\textwidth]{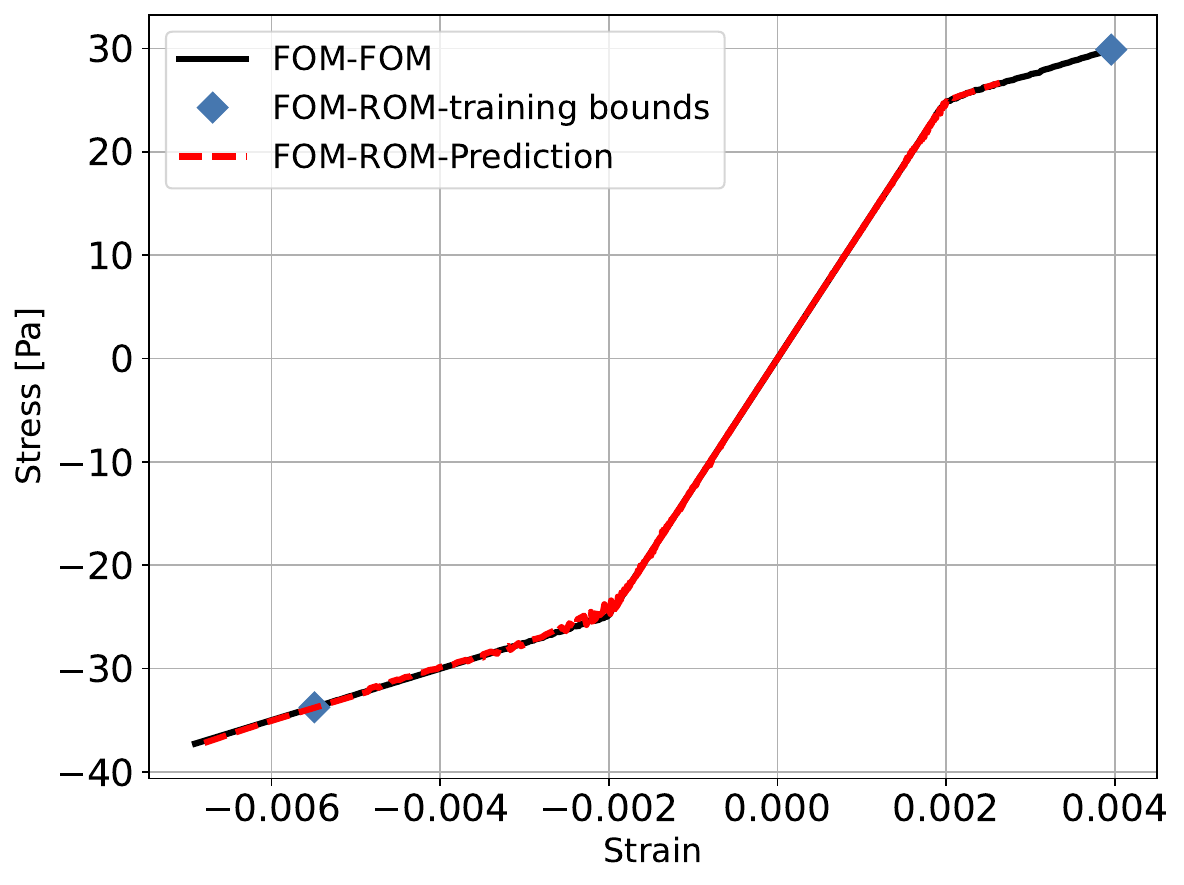}
\end{center}
\caption{Nonlinear elastic Stress-Strain law used for the vessel tube. The reconstructed curve from the ROM-FOM prediction is plotted along with the FOM-FOM model. We can also see the data points from the FOM-FOM simulation used for the ROM training.}
\label{fig:strstrain}
\end{figure}

After the ROM has been trained, we test the ROM-FOM coupling on the future prediction with $T_f = 120$~s at the same inlet velocity. We then choose another parameter vector $\hat{\pmb{\mu}} =  (0.9\,\,\,,4)^T$ without retraining the ROM, with $T_i = 0.1$~s We can see the difference between the two signals in Figure \ref{inlet_v}. The results for theses two cases are shown in Figures \ref{test1} and \ref{test2} respectively, compared to the HF FOM-FOM solution. We can see a significant accuracy achieved by the suggested ROM-FOM coupling. Moreover, the stress-strain law reconstructed using the ROM-FOM simulation is plotted in Figure \ref{fig:strstrain}, showing that the proposed ROM approach can capture the solid problem nonlinearities. Remarkably, we can also see that the ROM-FOM successfully predicted the vessel response in an extrapolated region in the strain response (strain region $\epsilon < -0.0055$) even for this nonlinear constitutive law, although it should be noted that it only involves a linear extrapolation from the phase space seen in the training.

\begin{figure}
\begin{center}
% \setlength{\unitlength}{1mm}
% \begin{picture}(210, 80)\label{iters_tube}
% {\framebox{\epsfig{file=Images/TOMAC_iters_tube_, scale=0.61,}}}
% \end{picture}
\includegraphics[width=.65\textwidth]{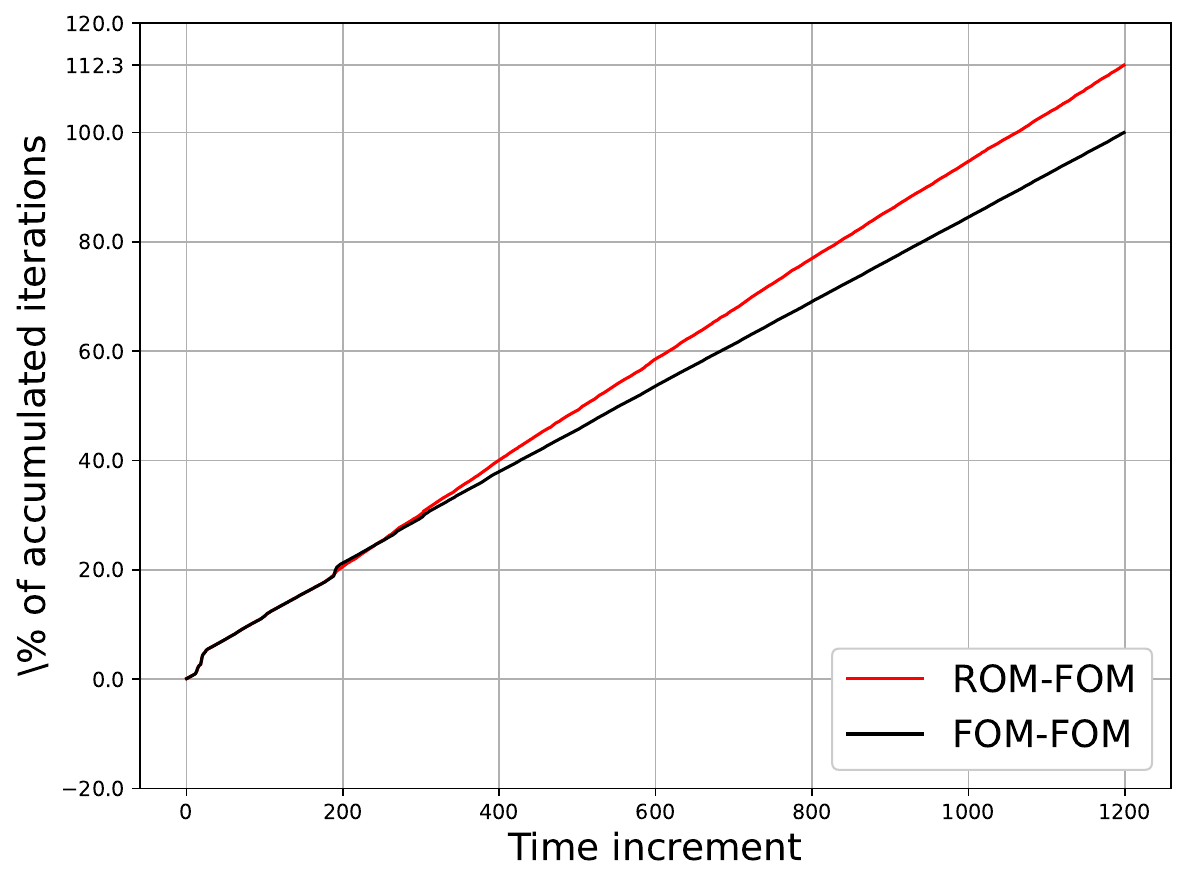}
\end{center}
\caption{Percentage of the accumulated number of iterations comparing the ROM-FOM to the FOM-FOM.}
\label{iters_stability}
\end{figure}

In addition, another important property of ROM-FOM coupling schemes is maintaining stability. This directly affects the final ROM speedup, since more instabilities result in an increased number of fixed-point iterations (or even a complete divergence), increasing therefore the overall computational cost. We show in Figure \ref{iters_stability} a comparison between the number of iterations until convergence at each time step for both the FOM-FOM and the ROM-FOM models. We observe -as expected- a slight increase in the number of iterations, due to the inaccuracy intrinsic to the reduced model. However, the ROM is accurate enough so that this increase remained limited (approximately $12$~\% more iterations at the end) and would not affect the overall speedup in more realistic cases.

The speedup achieved on the structural part for this test case is $\sigma \approx 96$. Nonetheless, there is no expected overall speedup since the fluid model is much more computationally costly than the solid FOM model ($\frac{T_f}{T_s}) \gg 1$. This is however merely a toy problem to evaluate the ROM accuracy and more realistic speedups will be shown in the following example.  

\subsection{Example 2: elastic flaps behind a cylinder wake}
%--------------------------------------------------
%
In this section, we consider a two-dimensional channel case, where an incompressible flow faces an elastic body with two mounted flap behind a rigid cylinder. We put $\rho = 1000\ kg/m^3$, $\nu_f = 0.001\ m^2/s$ and a fully developed Poiseuille inlet flow, with a maximum velocity of $v_{max} = 2.5\ m/s$ starting from $v = 0\ m/s$ at $t = 0s$ and increasing linearly until reaching $v_{max}$ at $t = 1 s$. This corresponds to a Reynolds number of $Re = 250$. The boundary condition at the top and bottom walls is a no slip condition, and a homogeneous Neumann boundary condition on the right boundary. A sketch of the test case is given in Figure \ref{testcase1}. The solid is governed by a hyperelastic constitutive law, using the Neo-Hookean model, where the stored elastic energy is
\begin{equation}
    W(\pmb{F}_s) = \frac{\mu_s}{2}\, (|\pmb{F}_s|^2 - 3 - 2\, \log(J))
\end{equation}
where we define $\mu_s = \dfrac{E}{2 (1 + \nu_s)}$ and we choose $E = 10 \times 10^6$~Pa and $\nu_s = 0.3$.

The fluid problem is discretised using triangular stabilized finite elements \cite{codina_stabilized_2001} and quadrilateral finite elements are used for the structural problem, with 4 ($X$ and $Y$) displacement degrees of freedom at element nodes. \texttt{KratosMutiphysics} \cite{dadvand_object-oriented_2010} was used as the finite elements software for both problems, using the Kratos modules named \texttt{FluidDynamicsApplication} and \texttt{StructuralMechanicsApplication} in a partitioned coupling. The fluid time step is 
$dt = 0.008$~s. The solid mesh consists of $1805$ nodes with $265$ nodes at the interface, giving $N = 530$ and $N_u = 3610$.

\begin{figure}
\begin{center}
\includegraphics[width=1\textwidth]{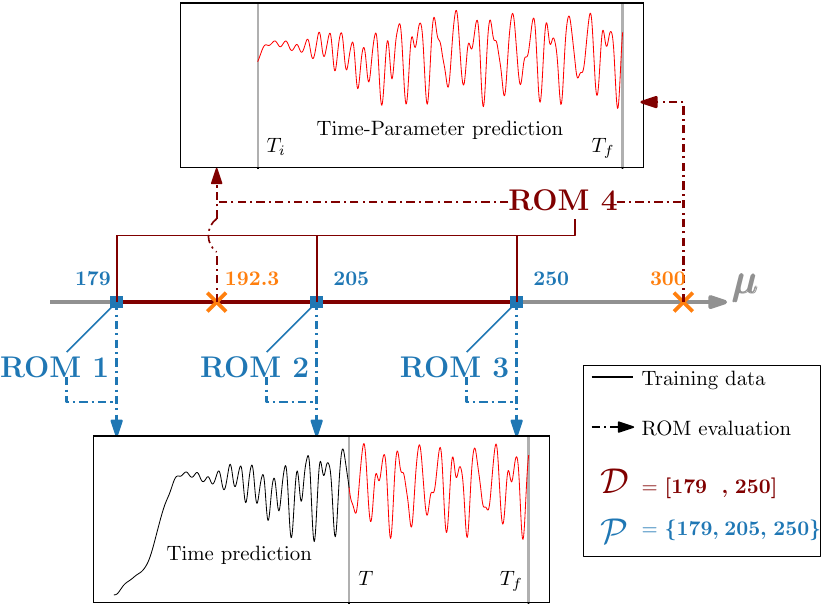}
\end{center}
\caption{Illustration of trained models and their evaluation strategy.}
\label{eval_illustr}
\end{figure}

As explained in section \ref{eval_st}, the Reynolds number $Re$ will be considered as a parameter $\mu$ in this test case. The training set will include 3 points $\mathcal{P} = \{ 178.6, 208.3, 250\}$, i.e $\mathcal{D} = [178.6, 250]$, and a time domain until $T = 3$~s.
In addition to time-predictions up to $T_f = 6$~s, two unseen parameters will be considered, $\hat{\mu}_1 = 192$ $\in \mathcal{D}$ and $\hat{\mu}_2 = 300$ $\not \in \mathcal{D}$, with $T_i = 0.8$~s. Accordingly, 4 different reduced order models will be assessed: 3 models in a time-prediction setting for each $\mu \in \mathcal{P}$ (hereinafter called \textbf{ROM 1}, \textbf{ROM 2} and \textbf{ROM 3} respectively), and the remaining ROM for time-parameter prediction for $\hat{\mu}_1$ and $\hat{\mu}_2$ (hereinafter called \textbf{ROM 4}). A summary of the ROM evaluation strategy is presented in Figure \ref{eval_illustr}.

\begin{figure}
\begin{center}
\includegraphics[width=1\textwidth]{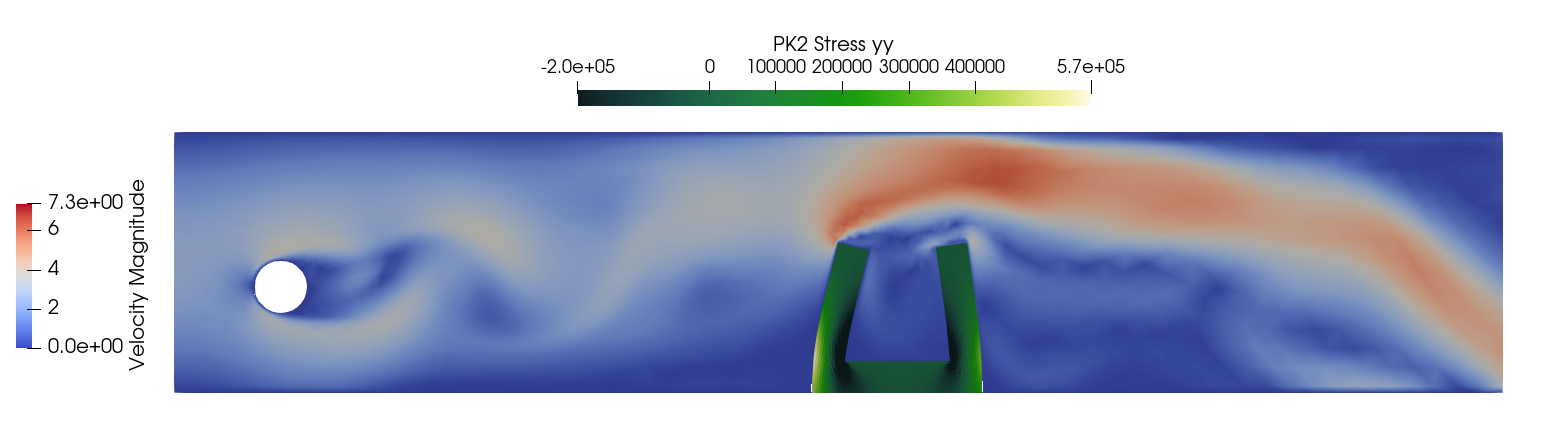}
\end{center}
\caption{Problem solution at $Re = 300$ and $t = 3.62$ s. The velocity magnitude is plotted in the fluid domain, while the $yy$ component of the second Piola Kirchoff stress (PK2) is showed in the solid domain.}
\label{contour}
\end{figure}

A total of $6890$ snapshots are used for training corresponding to $2222$, $2311$ and $2357$ snapshots for simulations at $Re$ $\in$ $\{ 178.6, 208.3, 250\}$ respectively. In Figure \ref{contour}, an example of the problem solution at $Re = 300$ and $t = 3.62$ s is reported for illustration.

\begin{figure}
\begin{center}
% \setlength{\unitlength}{1mm}
% \begin{picture}(160, 33)
% \put(   0,  0){\framebox{\epsfig{file=Images/wake_DoubleFlap-eps-converted-to.pdf, scale=1.,}}}
% \end{picture}
\includegraphics[width=1\textwidth]{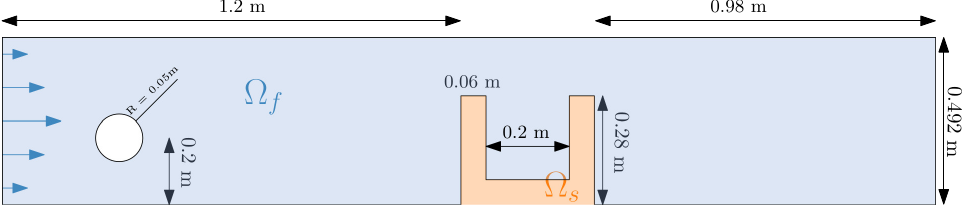}
\end{center}
\caption{Test case schematic explanation and dimensions.}
\label{testcase2}
\end{figure}

\subsubsection{Dimensionality Reduction}\label{case:DimReduc}
\begin{figure}
    \centering
    \includegraphics[width=.6\textwidth]{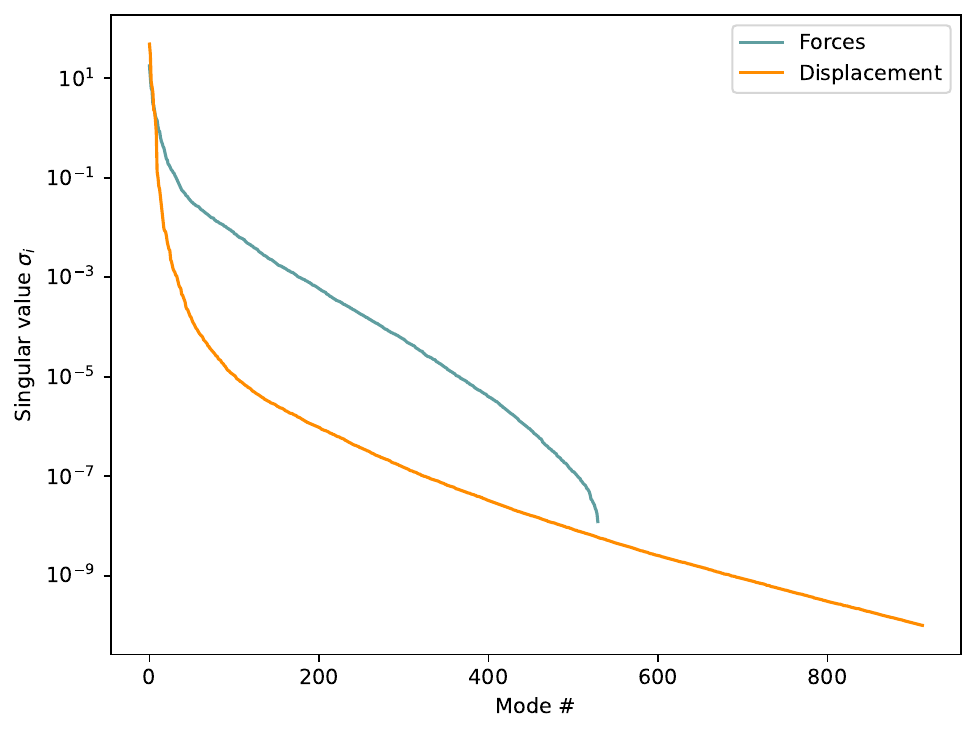}
    \caption{Singular values decay for the displacement and forces snapshots}
    \label{svd_Flap}
\end{figure}

In this case, the forces field on the structural interface has many irregularities that make the linear compression challenging. In Figure \ref{svd_Flap}, we show the energy quantity in the displacement and force fields modes through the singular values of the snapshots matrices. The decay for the forces field is much slower than the one of the displacement field. In order to determine the optimal number of modes used, we use $9$ displacement modes corresponding to $99.99\%$ of the energy, due to the observed fast decay of the singular values. For the forces field, we suggest a cross-validation strategy similar to the one used in the previous test case: $5\%$ of the training data are sampled for testing, and while varying the number of forces modes, the final error of the structural ROM $\varepsilon$ is evaluated as
\begin{equation}
    \varepsilon = \left\| \pmb{U}_{test} - ROM(\pmb{F}_{test}) \right\|_2
\end{equation}
Additionally, in order to evaluate the effect of the number of modes on the regression method accuracy, the error of the regression result on the training data $\varepsilon_{regression}$ is also evaluated as
\begin{equation}
    \varepsilon_{regression} = || \Tilde{\pmb{u}} - \mathcal{I}(\Tilde{\pmb{f}}) ||_2
\end{equation}
More details about the regression method used in this test case will be given in the next section. We conservatively choose to pick the maximum of modes between the two criteria. The result of the cross validation is shown in Figure \ref{crossV_full} for \textbf{ROM 4}, while the results for the other ROMs are reported in the Appendix. A clear threshold of the accuracy is reached after 45 modes. Indeed, there is a large gap between this choice and the energy-based criteria that results in 36 modes, which shows, again, the importance of this cross-validation step for more optimal choice of hyperparameters.

\begin{figure}
    \centering
    \includegraphics[width = .8\textwidth]{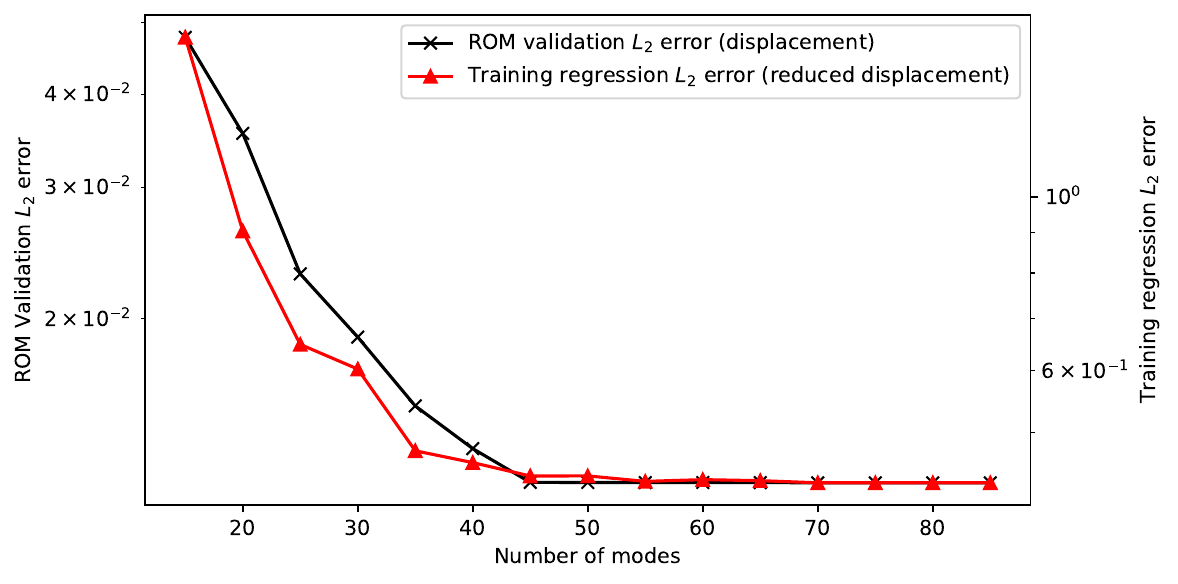}
    \caption{Testing error of the ROM results $\epsilon$ and the regression error on the training data $\epsilon_{regression}$ depending on the number of forces modes. Case of the parametric ROM 4.}
    \label{crossV_full}
\end{figure}

\subsubsection{Regression method}

For this test case, a polynomial regression of degree 2 is used. The Force-Displacement relationship is thus modeled as a second order polynomial:
\begin{equation}\label{regr_poly}
    \Tilde{\pmb{u}} = \pmb{W} [\Tilde{\pmb{f}} \Tilde{\otimes}  \Tilde{\pmb{f}}]
\end{equation}
where $\Tilde{\otimes}$ is a modified Kronecker product (now different than the product defined in (\ref{quad_recons}) because this product contains the linear terms as well and the intercept term $\pmb{\mathds{1}}$). The polynomial coefficients are arranged in $\pmb{W} \in \mathbb{R}^{r_u \times \hat{r}_f }$, where $\hat{r}_f = (r_f+1)(r_f+2)/2$ meaning that it scales as $\mathcal{O}(r_f^2)$.
We recall that during the cross-validation step $r_f$ was chosen as $r_f = 45$. This means that the number of polynomial coefficients is very large ($\hat{r}_f =1081$), and it is highly unlikely that all the polynomial terms are important for modeling $\mathcal{I(\cdot)}$. We thus propose using the Lasso regularization in order to obtain a parsimonious model with as fewest terms as possible, the minimization is written as:
\begin{equation}
    \pmb{W}_i = arg \min_{\hat{W}_i} ||\Tilde{\pmb{u}}_i - \sum_{j=1}^{\hat{r}_f} \hat{W}_{ij} [\Tilde{\pmb{f}} \otimes  \Tilde{\pmb{f}}]_j ||_2  + \lambda \sum_{j=1}^{\hat{r}_f} |\hat{W}_{ij}| \qquad
\forall \ i \in \{1 \cdots r_u\}
\end{equation}
where $\pmb{W}_i$ and $\pmb{u}_i$ are the $i^{th}$ rows and $[\Tilde{\pmb{f}} \otimes  \Tilde{\pmb{f}}]_j$ is the $j{th}$ row of the matrices defined in (\ref{regr_poly}).
The parameter $\lambda$ promotes the sparsity of the solution $\pmb{W}$ and usually requires fine-tuning. One approach to determine $\lambda$ is to use model selection criteria combined with the LARS algorithm \cite{LASSO} for finding a Lasso solution. This allows the generation of a path of Lasso solutions and the evaluation of criteria, namely the Bayesian information criterion (BIC) along this path. The regularization coefficient $\lambda$ corresponding to the minimum of BIC will be chosen. We use the \texttt{scikit-learn}\cite{sktlrn} implementation of this method.

\begin{figure}
    \centering
    \includegraphics[width = \textwidth]{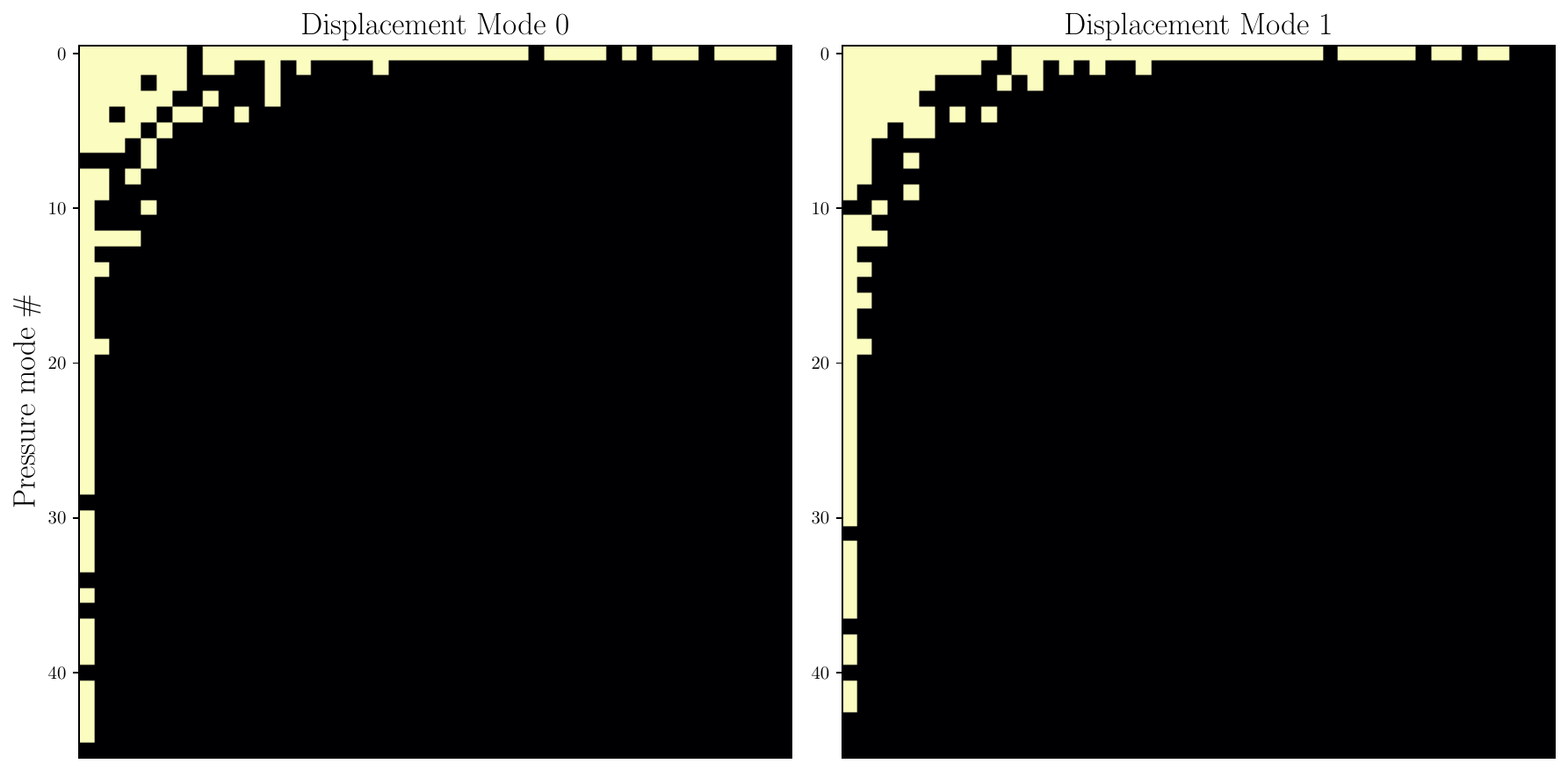}
    \caption{Illustration of the non-zero terms (yellow pixels) among all the $2^{nd}$ degree polynomial terms. Each row/column represents a pressure mode so that the $ij$ pixel represents the term corresponding to the product of $i^{th}$ mode with $j^{th}$ mode, while the mode $0$ represents the term $1$. Case of \textbf{ROM 4} and the first two displacement modes.}
    \label{modes_lasso_full}
\end{figure}

We show in Figure \ref{modes_lasso_full} the non-zero terms as a result of the Lasso optimization. We only show the terms associated with the first two outputs (the first two displacement modes) for the model \textbf{ROM 4}. For brevity, the other figures associated with the other ROMs are reported in the Appendix. We can clearly see that only few terms remain after the Lasso optimization. In addition, two important observations are to be made here, the Lasso optimization confirms the necessity of using as much as 45 modes, since many terms associated with these modes are considered. Second, terms associated with interacting modes are also important, showing that modeling the nonlinearity of the force-displacement relationship as a second-degree polynomial function is indeed an adequate choice.

\subsubsection{ROM-FOM coupling accuracy and speedup}

\begin{figure}
\begin{center}
\includegraphics[width=1\textwidth]{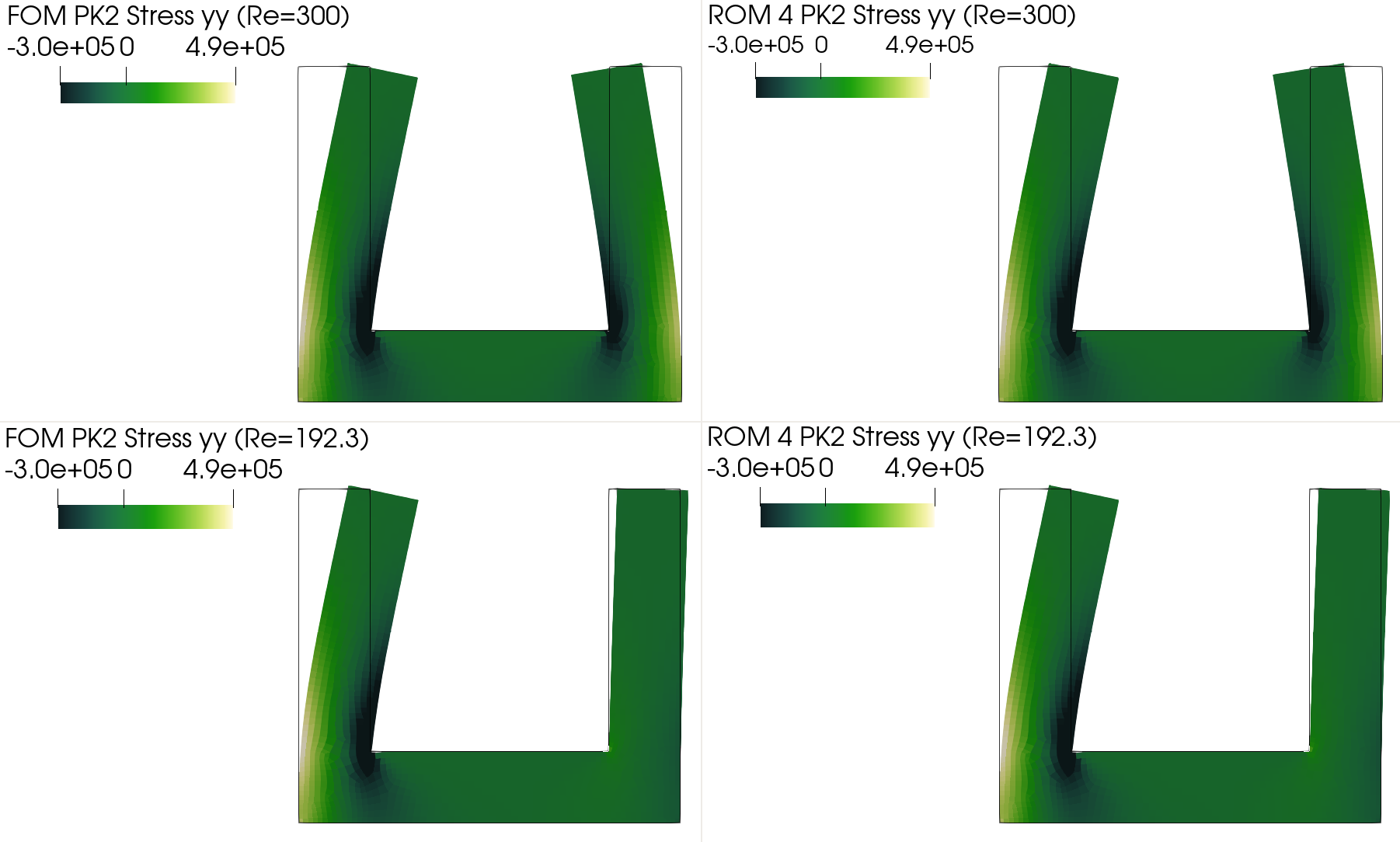}
\end{center}
\caption{The ROM-FOM (using \textbf{ROM 4}) results compared to the FOM-FOM results. This is taken at $t = 5.84$ s. The $yy$ component of the PK2 stress field is showed while also showing the complete solid deformation.}
\label{fig:deform}
\end{figure}

\begin{figure}
\begin{center}
\includegraphics[width=1\textwidth]{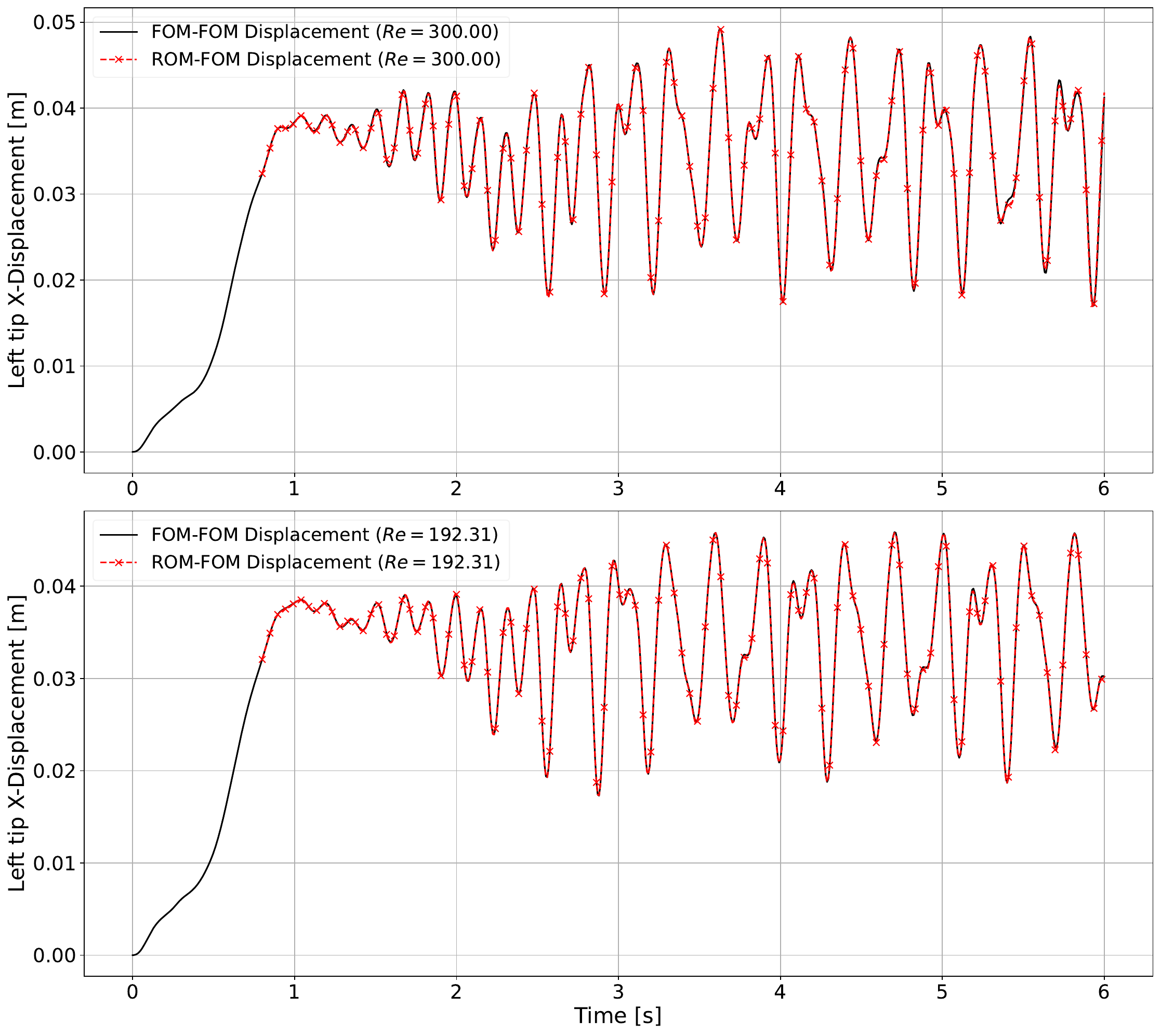}
\end{center}
\caption{The ROM-FOM left tip x-displacement of the solid body using \textbf{ROM 4} (evaluated at $Re = 192$ and $Re = 300$), compared to the FOM-FOM solution.}
\label{leftTipDisplacement}
\end{figure}

\begin{figure}
\begin{center}
\includegraphics[width=1\textwidth]{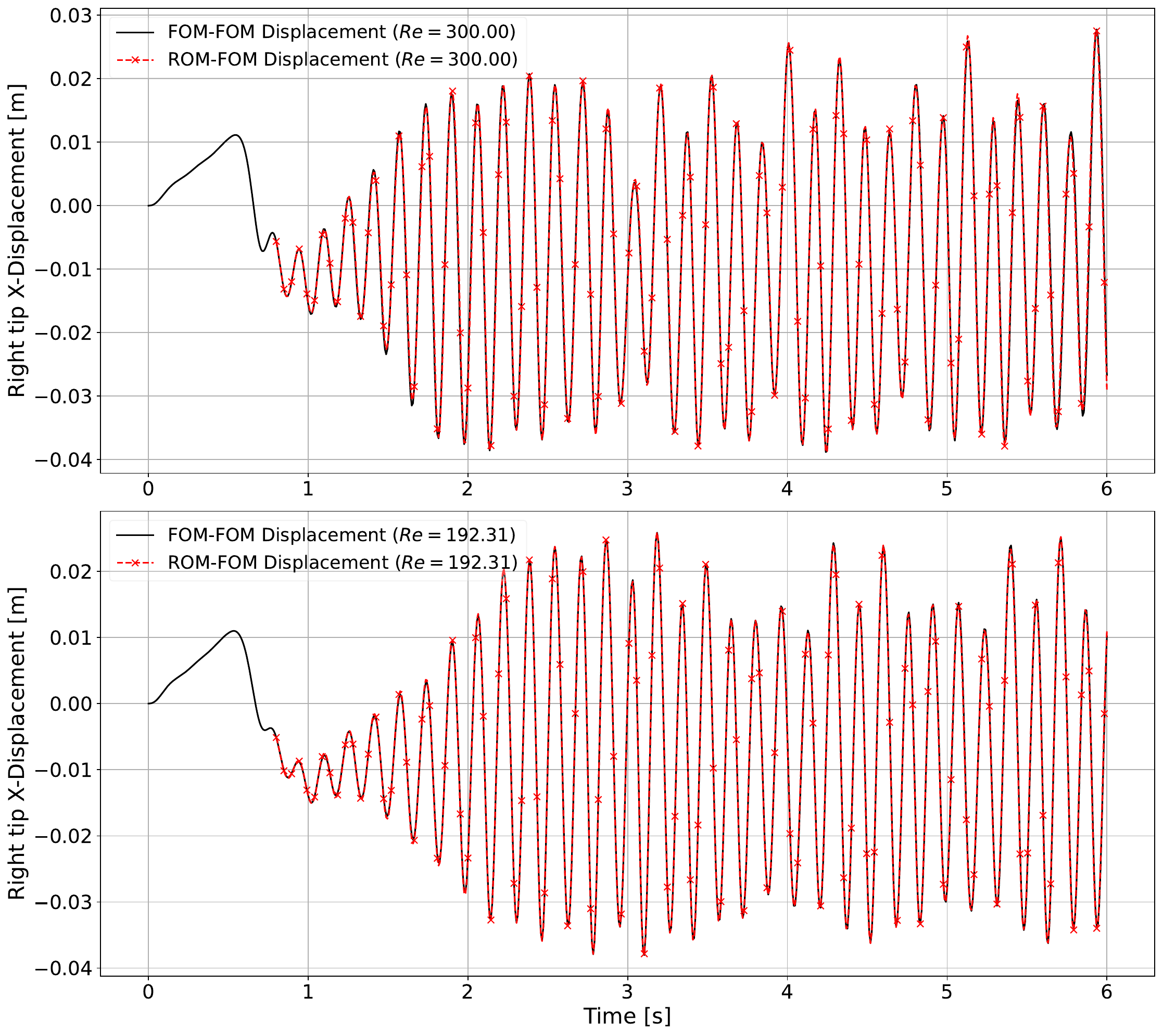}
\end{center}
\caption{The ROM-FOM right tip x-displacement of the solid body using \textbf{ROM 4} (evaluated at $Re = 192$ and $Re = 300$), compared to the FOM-FOM solution.}
\label{rightTipDisplacement}
\end{figure}

\begin{figure}
\begin{center}
\hspace*{-1.7cm}\includegraphics[width=1.3\textwidth]{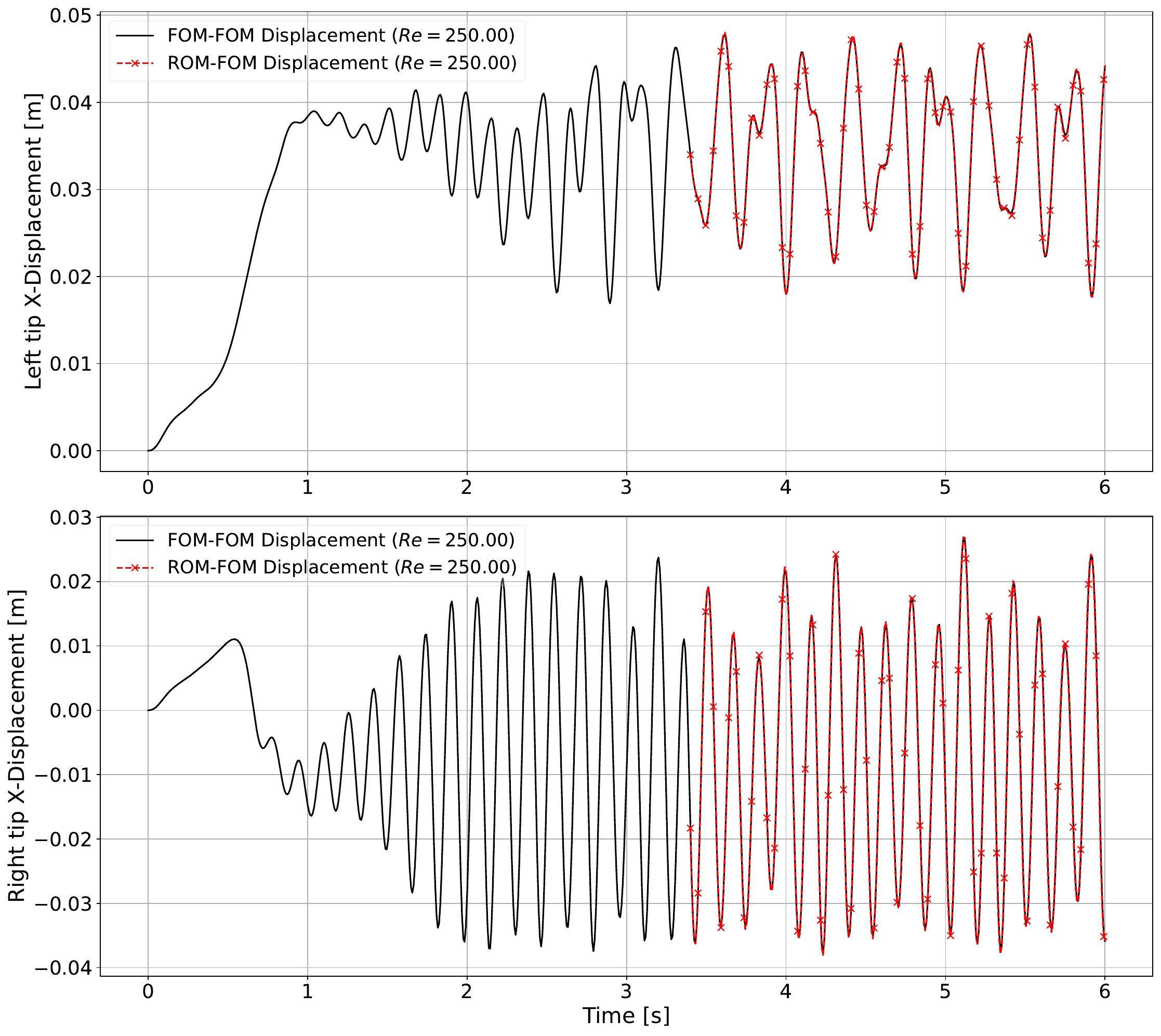}
\end{center}
\caption{The ROM-FOM left tip x-displacement of the solid body using \textbf{ROM 3} compared to the FOM-FOM solution.}
\label{TipDisplacementPred}
\end{figure}

\begin{figure}
\begin{center}
\includegraphics[width=1\textwidth]{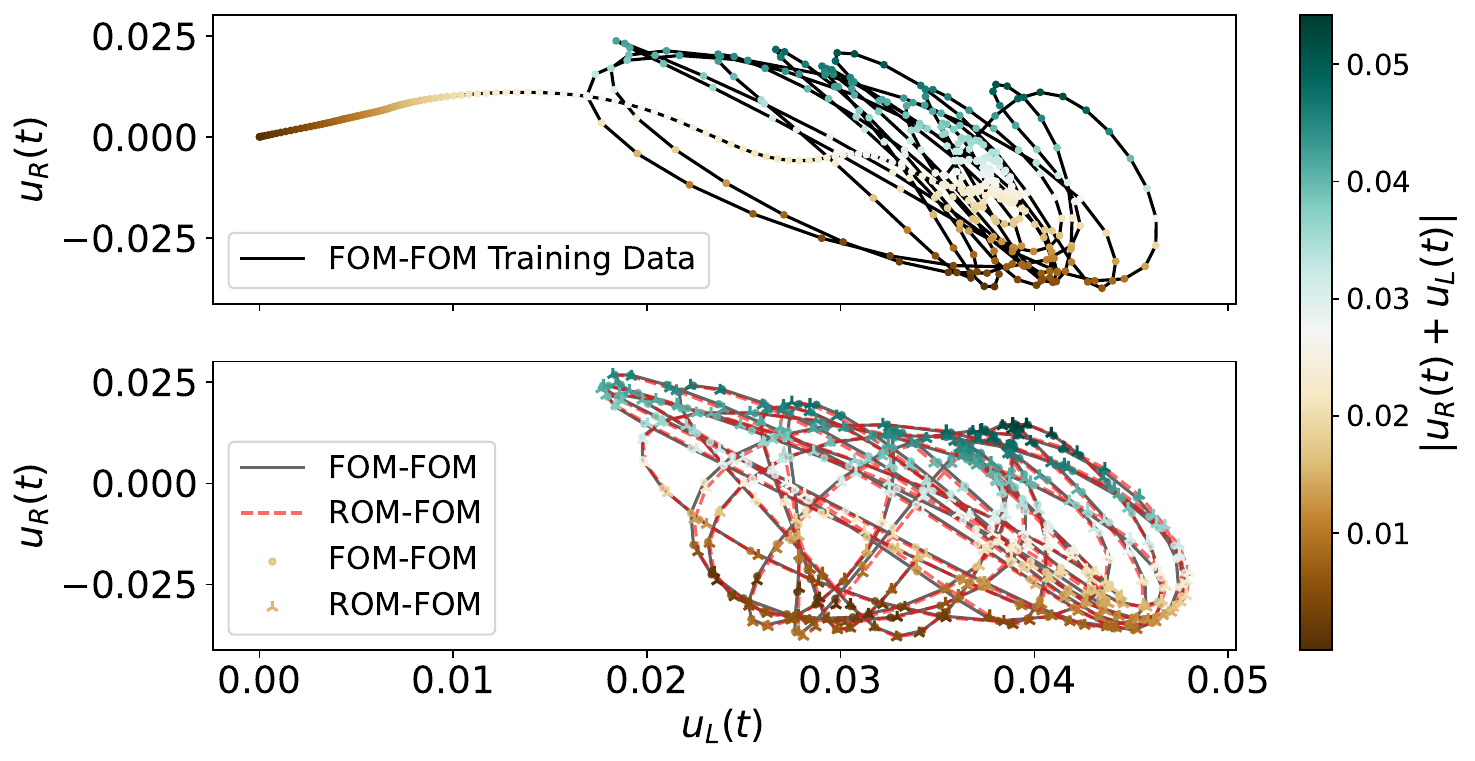}
\end{center}
\caption{The X-displacement phase space ($u_L(t)$, $u_R(t)$) where $u_L(t)$ is the left-most node displacement and $u_L(t)$ is the right-most node displacement. \textbf{Top}: Trajectory seen in training ($Re = 250$). \textbf{Bottom}: A comparison between the FOM-FOM (black line) and ROM-FOM (red dashed line) results. Case of \textbf{ROM 3}}
\label{phase-space_10}
\end{figure}

\begin{figure}
\begin{center}
\includegraphics[width=1\textwidth]{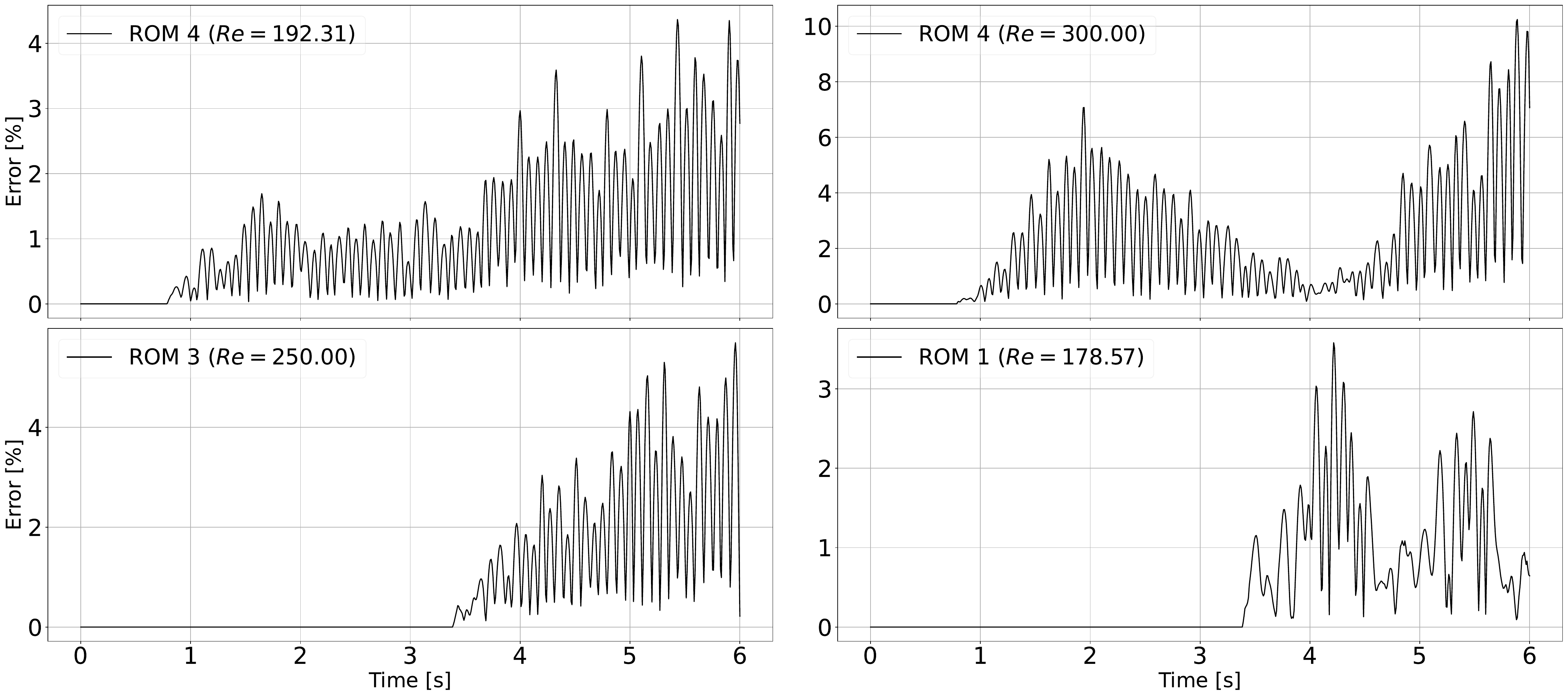}
\end{center}
\caption{The ROM-FOM $L_2$ error $e(t)$ as defined in (\ref{double_flap_error}) for the previous cases.}
\label{full_error_L2}
\end{figure}

The \textbf{ROM 4} evaluation results are shown in Figure \ref{leftTipDisplacement} where we show the time evolution of the left-most solid tip $x$-displacement, and similarly for the right-most tip in \ref{rightTipDisplacement}. We also show the same results in a time-prediction setting using \textbf{ROM 1} in Figure \ref{TipDisplacementPred}. The complete deformation with a map of the PK2 stress field is shown for both the FOM and ROM results in Figure \ref{fig:deform}. The relative $L_2$ error on the complete displacement field defined as 
\begin{equation}\label{double_flap_error}
    e(t) = \frac{||\pmb{U}(t) - \hat{\pmb{U}}(t)||_2}{ <{||\pmb{U}(t)||_2}>}
\end{equation}
is also shown in Figure \ref{full_error_L2}. A highly accurate prediction of the ROM-FOM model is obtained for each case. For the unseen parameter outside the training space $\mathcal{D}$ ($Re = 300)$, the error is larger, but remains under 10$\%$ for the considered time region. Remarkably, in the case of \textbf{ROM 1}, \textbf{ROM 2} and \textbf{ROM 3}, where the training was done on a single parameter and where the evaluation was done on an unseen time range, only $4\%$ of error is observed, despite the complex dynamics. To further highlight this, we show the displacement phase-space ($u_L(t)$, $u_R(t)$) in Figure \ref{phase-space_10},  where $u_L(t)$ is the left-most node displacement and $u_R(t)$ is the right-most node displacement. More precisely, we show the space seen in training (Top) and the one seen in the testing phase (Bottom) in the case of \textbf{ROM 3}. This figure demonstrates a good agreement between the FOM-FOM and ROM-FOM results. Again, the complexity of the dynamics is well handled since a high-fidelity FOM is kept on the fluid part. We argue that as long as the amplitudes of the fluid forces are close to those seen in training, the proposed ROM strategy will enable a good prediction accuracy, even when extrapolating in time and parametric space, at least when the flow regime remains the same.

\begin{figure}
\begin{center}
\includegraphics[width=.7\textwidth]{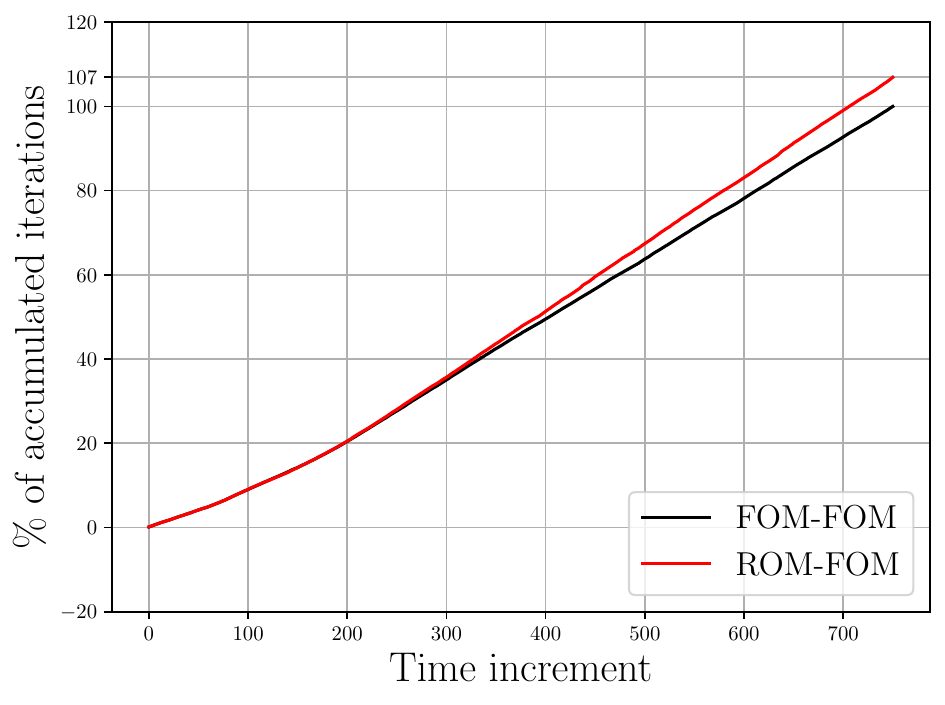}
\end{center}
\caption{Accumulated number of iterations in a comparison between the ROM-FOM and FOM-FOM results. Case of \textbf{ROM 4} evaluated at $Re = 300$}
\label{fig:iterations_doubleflap_full}
\end{figure}

As an indicator of the ROM-FOM stability, in Figure \ref{fig:iterations_doubleflap_full}, we show the number of coupling iterations needed to convergence, compared to the FOM-FOM results. We see only $7\%$ additional iterations in the case of \textbf{ROM 4} evaluation at $Re = 300$. The other cases are reported in the Appendix. Such a slight increase in the overall coupling iterations show the good stability of the ROM-FOM scheme, due to the high accuracy of the solid ROM.

\begin{figure}
\begin{center}
\includegraphics[width=.9\textwidth]{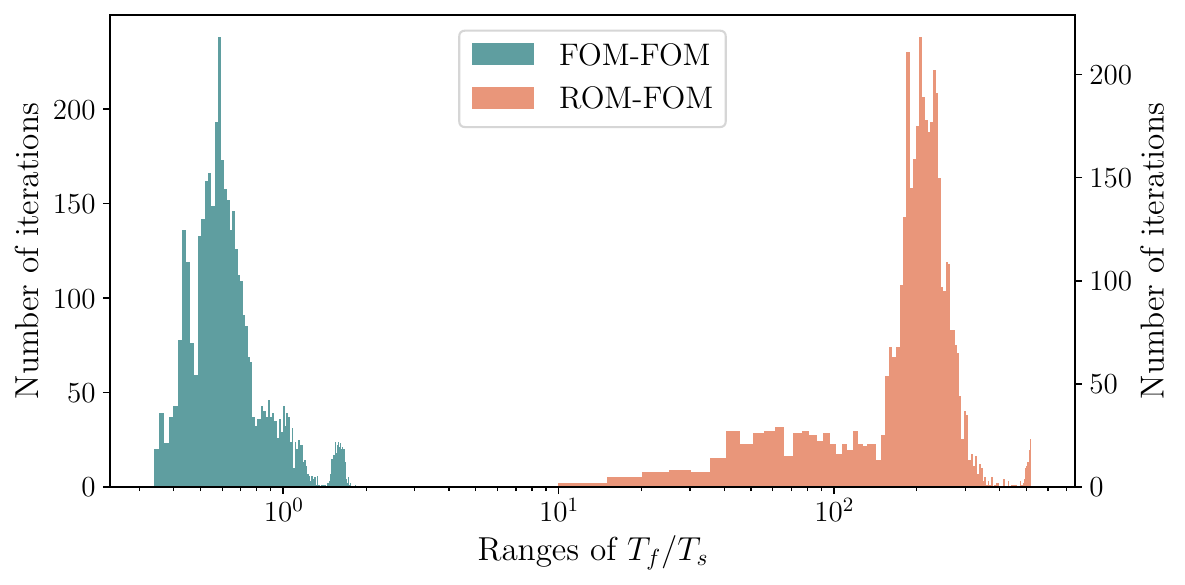}
\end{center}
\caption{Distribution of the ratios of CPU time $\frac{T_f}{T_s}$ for all subiterations for both the ROM-FOM to the FOM-FOM strategies.}
\label{solver_ratios_full}
\end{figure}

On an 8-Cores Mac M1 laptop, the CPU time of the ROM solid solver is on average (case of \textbf{ROM 4} at $Re = 300$) $0.0028\;s$ as opposed to $T_s = 0.724\;s$ for the solid FOM solver, making the speedup obtained for the solid problem $\sigma \approx 260$. The total CPU time for the complete FOM-FOM simulation is $5763\;s$  and $3181\;s$ for the ROM-FOM simulation, making the overall speedup for the FSI simulation around $s \approx 1.81$. We should note that the ratio $\frac{T_f}{T_s}$ varies during the simulation, since the internal solvers' convergence takes different number of nonlinear iterations depending on the simulation time steps. For a more detailed analysis, we show in Figure \ref{solver_ratios_full} the ratio $\frac{T_f}{T_s}$ for every subiterations and its distribution, while comparing both models, we can see the large difference when modifying the solid solver to make it much faster than the fluid solver. Average speedups and CPU time of training the ROMs are reported in the Appendix.

\bigskip

\section{Conclusions}\label{sect:concl}

We have presented a ROM-FOM coupling strategy designed to reduce the computational cost of partitioned FSI simulations, through the reduction of the solid sub-problem. The proposed approach is data-driven, can be implemented in a completely non-intrusive framework, and is well-suited for nonlinear elastic solid problems under quasi-static load. Using a dimensionality reduction of the fluid forces as well as the displacement field on linear subspaces, a regression model between the two fields is constructed in the latent space, followed by a reconstruction towards the high-dimensional space of the displacement field using a quadratic manifold approximation, for a reconstruction as much accurate as possible. The obtained ROM can accurately predict the full displacement field online with a significant speedup compared to the reference FOM-FOM solution. The performance of the proposed ROM strategy has been demonstrated on two different test cases where we have shown a significant accuracy of the ROM-FOM coupling even for complex dynamics tracking. Finally, the proposed methodology was shown to perform very well in both time and parametric extrapolated region space.
Despite the novelty of this type of ROM-FOM coupling scheme, and its significant potential for application on many common FSI problems, improvements can still be made to enhance the robustness of this method. For example, further understanding of the instability of the FSI simulations - a property intrinsic to the partitioned coupling schemes - will allow a better design of the structural ROMs, avoiding additional coupling iterations, and achieving better speedups. Nonlinear dimensionality reduction methods, especially for the fluid forces field, could also make the ROM less complex while maintaining good generalization properties. Additionally, multidimensional parametric spaces, including parameters of the solid problem itself (e.g material constitutive law parameters) will also be investigated in the future.
%Additionally, an extension for the structural ROMs to be suited for inelastic materials, would open the door for a vast applicability on various solid problems.

\section{Acknowledgements}
This work has been funded by the ANR (Agence Nationale de la Recherche), Altair Engineering and Michelin.

\appendix

\section{Appendix: Figures for detailed results of the constructed ROMs}
\subsection{\textbf{Choices of number of forces modes: }}
As explained in section \ref{case:DimReduc}, a cross-validation is done for every ROM training, in order to find the optimal latent dimension of the fluid forces subspace, the results for \textbf{ROM 4} having been showed in Figure \ref{crossV_full}, the results for the other ROMs will be shown in the following Figure \ref{fig:crossV_full_Appendix}.
These results and the energy-criterion for the displacement modes give the optimal choices that are summarized in Table \ref{table:modes} 
\begin{table}[ht]
\caption{Cross validation results.}
\begin{tabular}{ |p{3cm}||p{2cm}|p{2cm}|p{2cm}|p{2cm}|  }
 \hline
  & \textbf{ROM 1} &\textbf{ROM 2}&\textbf{ROM 3}&\textbf{ROM 4}\\
 \hline
 \# of modes of \pmb{F}   & 45    &40&   50&45\\
 \hline
 \# of modes of \pmb{U}   & \multicolumn{4}{|c|}{9}\\
 \hline
\end{tabular}
  \label{table:modes}
\end{table}
\begin{figure}
\begin{minipage}{.5\linewidth}
\centering
\subfloat[\textbf{ROM 1} case]{\label{main:a}\includegraphics[scale=.4]{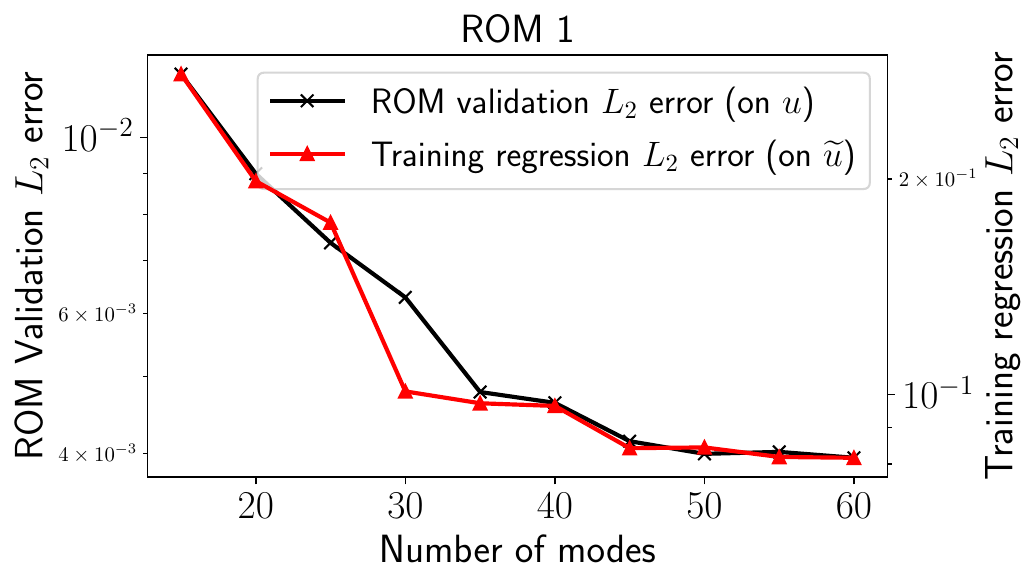}}
\end{minipage}%
\begin{minipage}{.5\linewidth}
\centering
\subfloat[\textbf{ROM 2} case]{\label{main:b}\includegraphics[scale=.4]{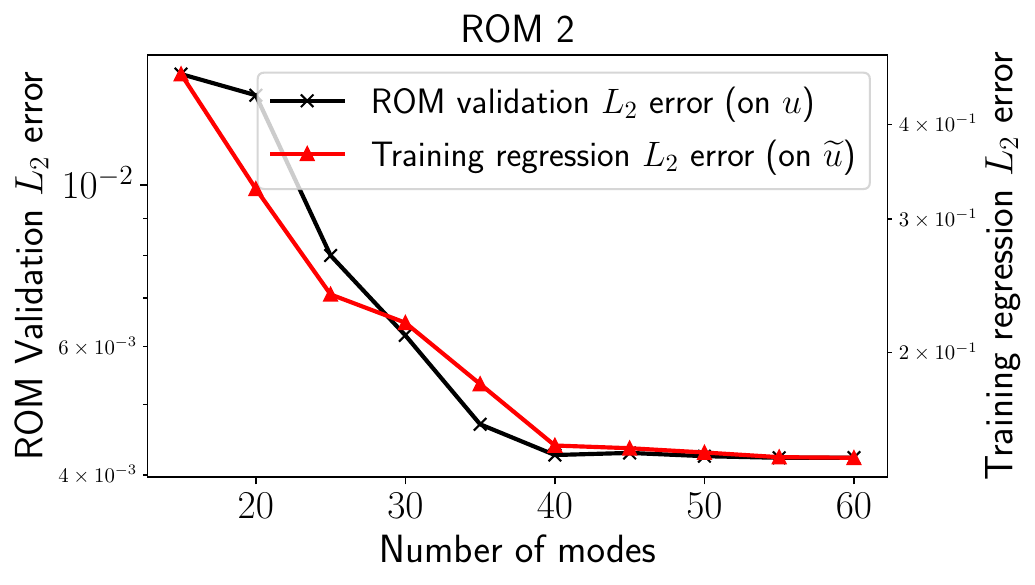}}
\end{minipage}\par\medskip
\centering
\subfloat[\textbf{ROM 3} case]{\label{main:c}\includegraphics[scale=.45]{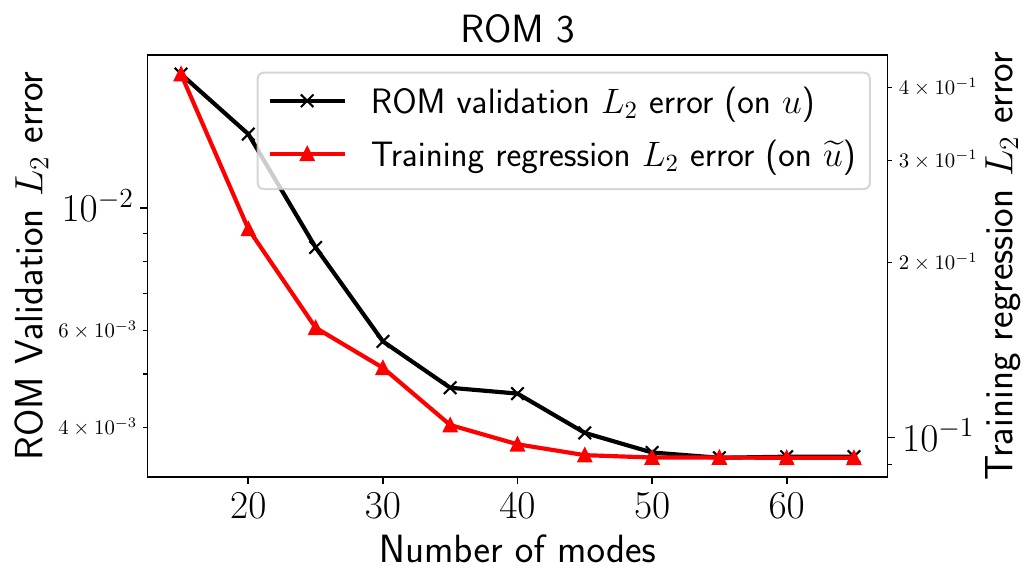}}
\caption{Testing error of the ROM results $\epsilon$ and $\epsilon_{regression}$ for cross-validation.}
\label{fig:crossV_full_Appendix}
\end{figure}
\subsection{\textbf{Polynomial terms after the Lasso optimization: }}
Similarly to Figure \ref{modes_lasso_full}, we show the equivalent results for the other trained ROMs.
\begin{figure}
    \centering
    \includegraphics[width =.9 \textwidth]{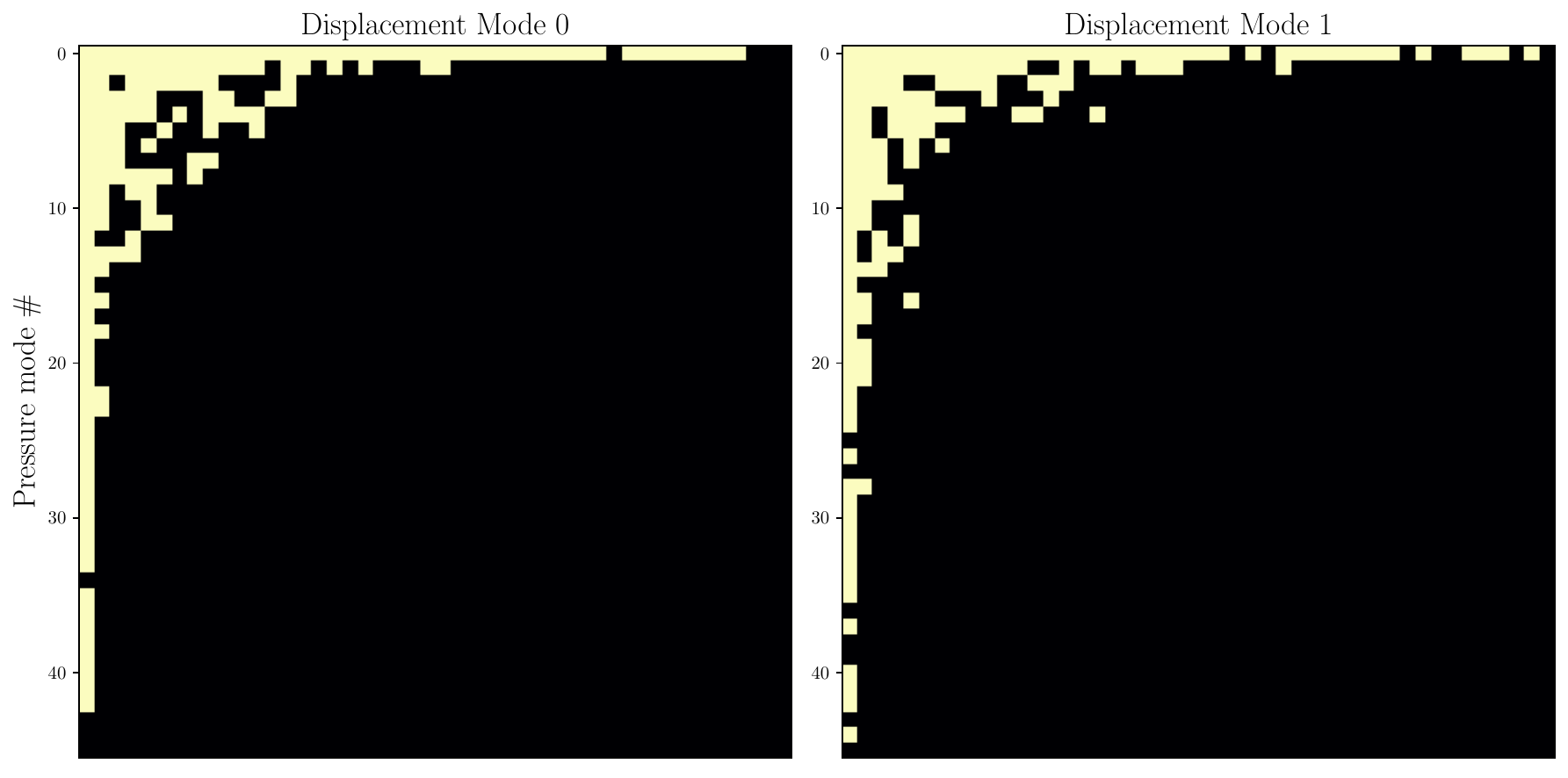}
    \caption{Non-zero polynomial terms. \textbf{ROM 1} }
    \label{modes_lasso_14}
\end{figure}
\begin{figure}
    \centering
    \includegraphics[width =.9 \textwidth]{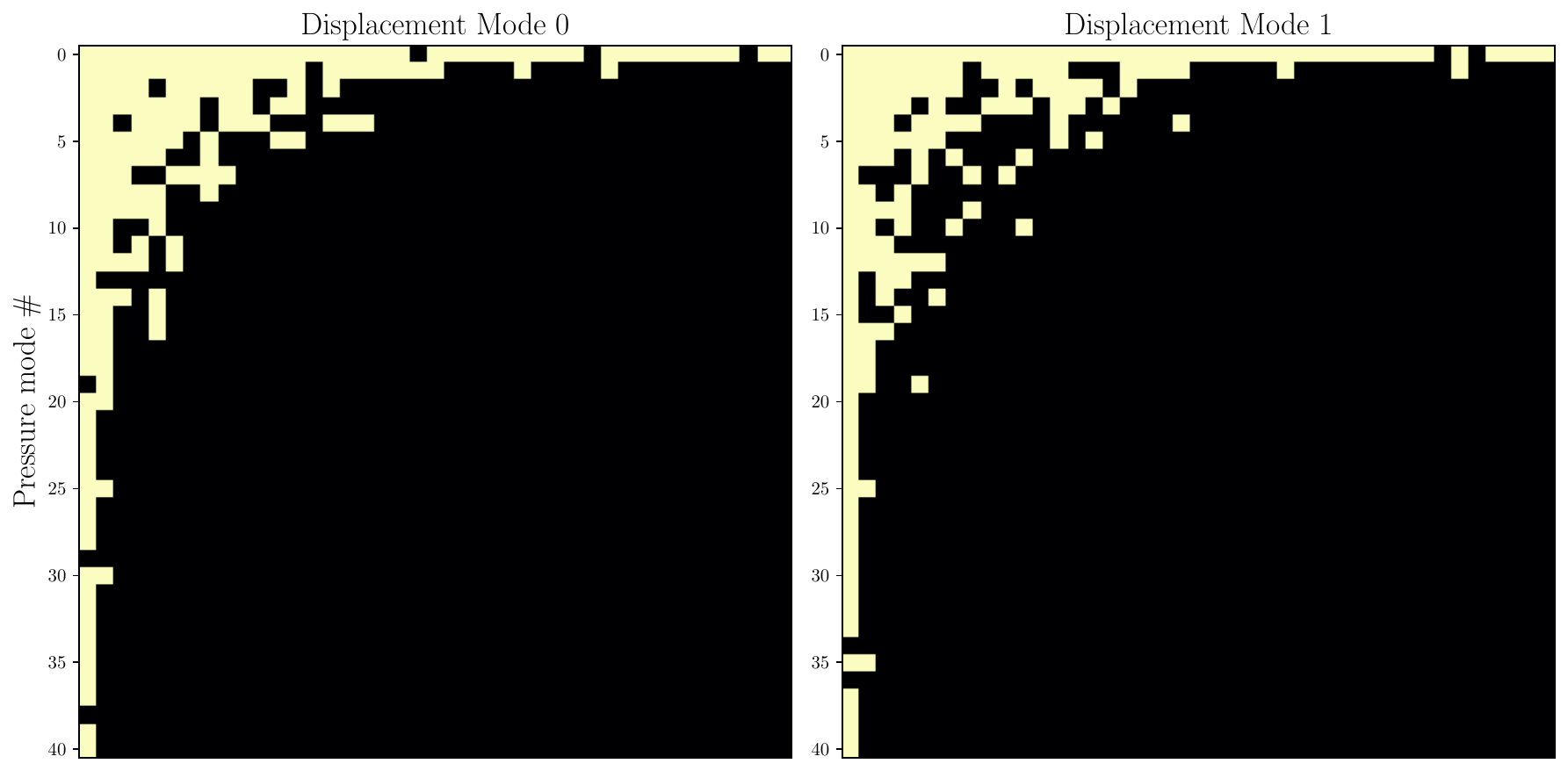}
    \caption{Non-zero polynomial terms. \textbf{ROM 2} }
    \label{modes_lasso_12}
\end{figure}
\begin{figure}[ht]
    \centering
    \includegraphics[width =.9 \textwidth]{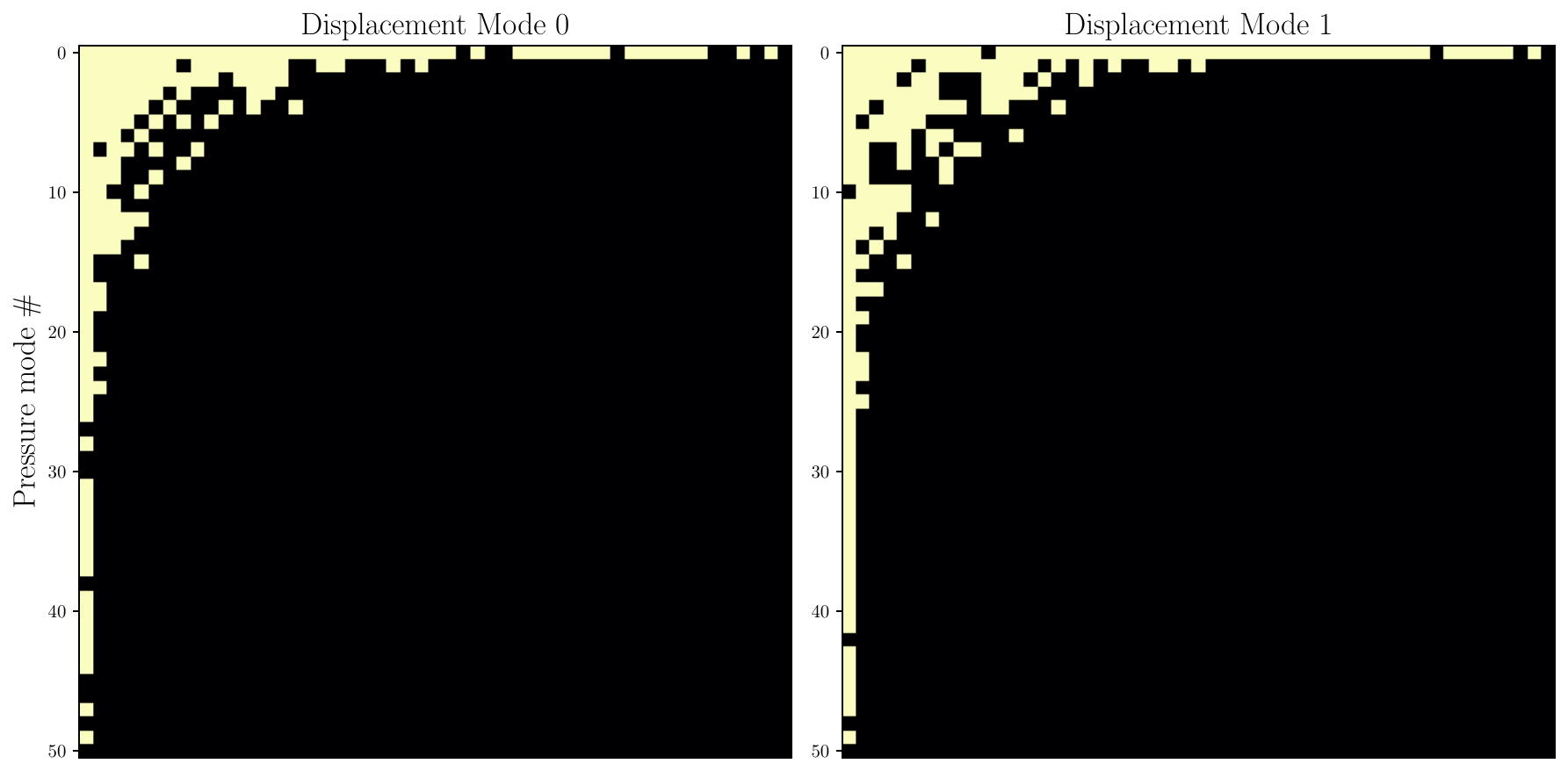}
    \caption{Non-zero polynomial terms. \textbf{ROM 3} }
    \label{modes_lasso_10}
\end{figure}
\subsection{\textbf{ROM evaluation: Displacement results: }}
We show in the following Figure \ref{fig:disp_comparison_2Reynolds} the results of the ROM-FOM and FOM-FOM comparisons in terms of displacement values at the tips of the structure.
\begin{figure}
\begin{center}
\hspace*{-1.7cm}\includegraphics[width=1.3\textwidth]{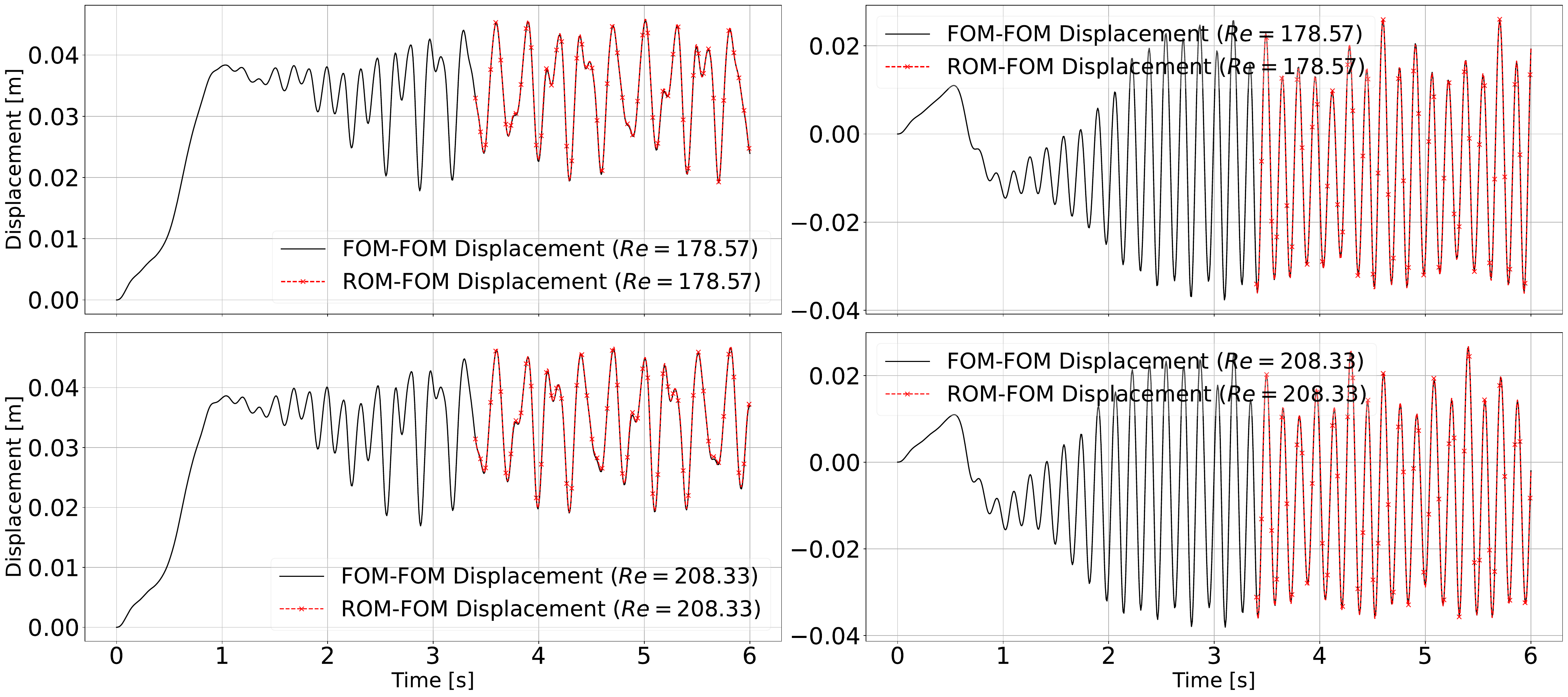}
\end{center}
\caption{The ROM-FOM tips x-displacement of the solid body compared to the FOM-FOM solution. \textbf{Top:} case of \textbf{ROM 1} - \textbf{Bottom:} case of \textbf{ROM 2}. \textbf{Left:} Left-most tip - \textbf{Right:} Right-most tip}
\label{fig:disp_comparison_2Reynolds}
\end{figure}
\subsection{\textbf{ROM evaluation: Coupling iterations: }}
\begin{figure}
\begin{center}
\includegraphics[width=.9\textwidth]{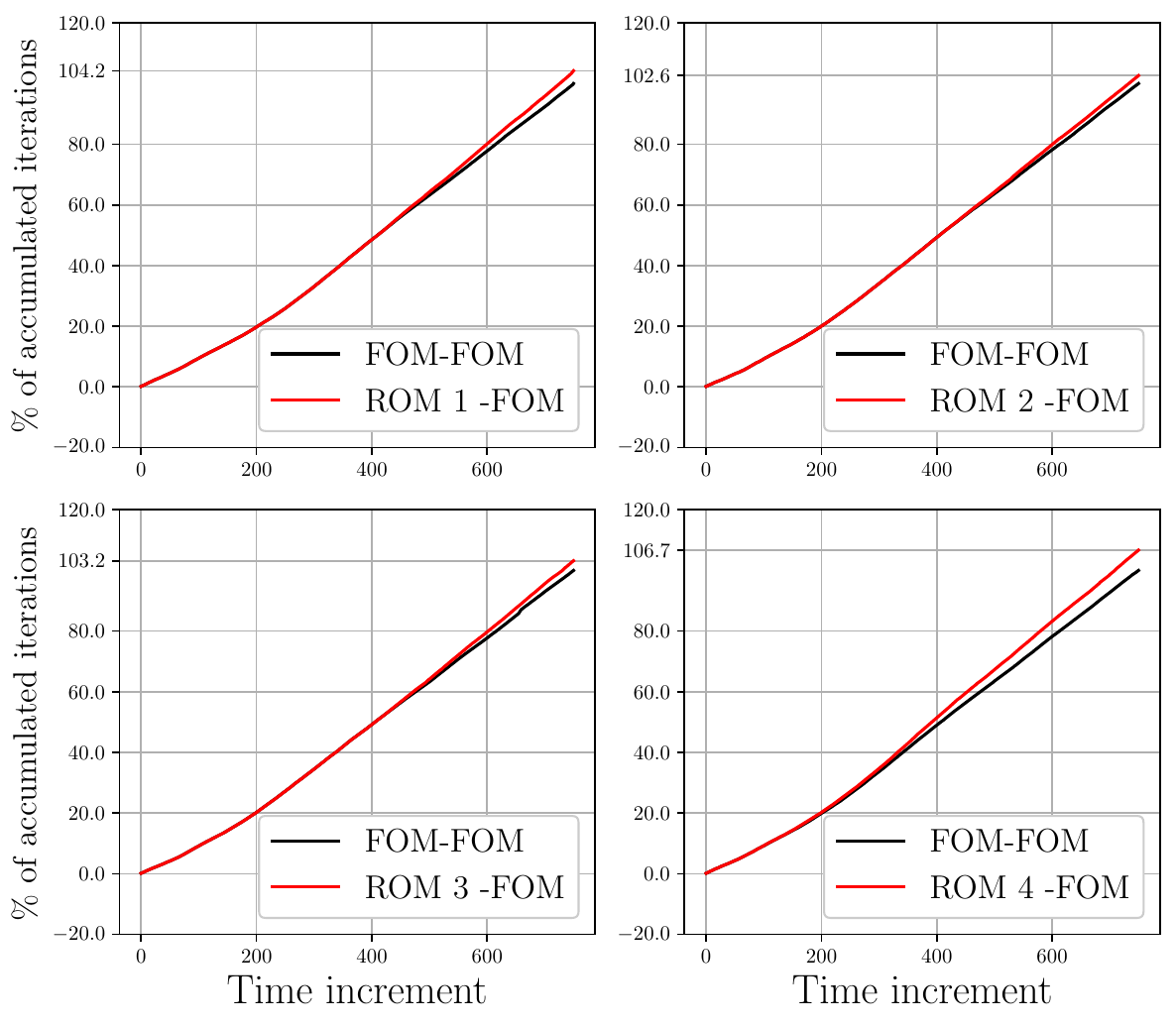}
\end{center}
\caption{Comparison of total fixed-point iterations between the FOM-FOM and ROM-FOM for \textbf{ROM 1}, \textbf{ROM 2}, \textbf{ROM 3} and \textbf{ROM 4} (at $Re = 192.3$)}
\label{fig:iters_doubleflap_other}
\end{figure}

Similarly to Figure \ref{fig:iterations_doubleflap_full}, we show the coupling iterations comparison for the other ROMs in Figure \ref{fig:iters_doubleflap_other}.
\subsection{\textbf{ROM evaluation: Speedups: }}
In this section, we show the average speedup - On an 8-Cores Mac M1 laptop with a $16$ GB RAM - achieved by the ROM, on the structural solver itself and on the total simulation. We note that the displacement field reconstruction (\ref{quad_recons}) is taken into account in the ROM computation time.
\begin{table}[ht]
\caption{Speedup results.}
\begin{tabular}{ |p{4cm}||p{1.5cm}|p{1.5cm}|p{1.5cm}|p{2.4cm}|p{2.4cm}| }
 \hline
  & \textbf{ROM 1} &\textbf{ROM 2}&\textbf{ROM 3}&\textbf{ROM 4} ($Re = 192.3$)&\textbf{ROM 4} ($Re = 300$)\\
 \hline
 Training time [s]  & 20.98    &22.08&   33.55& \multicolumn{2}{|c|}{52.77}\\
 \hline
 (Training+Cross validation) time [s]  & 181.9    &188.12&   242.12&\multicolumn{2}{|c|}{1348.58}\\
 \hline
 Average solid speedup  & 236.35    &183.64&   250.75&239.51&259.78\\
 \hline
 Total speedup  & 1.814    &1.88&   1.795&1.656&1.81\\
 \hline
\end{tabular}
  \label{table:speedups}
\end{table}


%%       \citet{<label>} ==> Jones et al. [21]
%%       \citep{<label>} ==> [21]
%%






% \end{thebibliography}

\clearpage

\bibliographystyle{elsarticle-num-names} 
\bibliography{elsarticle-template-num-names}

\end{document}